\documentclass[journal,twoside]{IEEEtran}
\usepackage{cite}

\usepackage{graphicx}
\graphicspath{{./figs/}}
\ifCLASSINFOpdf
\else
\fi
\usepackage{amsmath}
\usepackage{array}

\usepackage[caption=false,font=footnotesize]{subfig}
\usepackage{url}

\usepackage{siunitx}
\usepackage{xcolor}

\begin{document}
%
\title{Neuromodulation of Neuromorphic Circuits}
%
%
%

\author{Luka~Ribar
        and~Rodolphe~Sepulchre
\thanks{L. Ribar has received funding from Trinity College, Cambridge.}%
\thanks{The research leading to these results has received funding from the European Research Council under the Advanced ERC Grant Agreement Switchlet n.670645.}%
\thanks{L. Ribar and R. Sepulchre are with the Department of Engineering, University of Cambridge, CB2 1TZ Cambridge, U.K. (email: lr368@cam.ac.uk; r.sepulchre@eng.cam.ac.uk)}}

\maketitle

\begin{abstract}
We present a novel methodology to enable control of a neuromorphic circuit in close analogy with the physiological neuromodulation of a single neuron. The methodology is general in that it only relies on a parallel interconnection of elementary voltage-controlled current sources. In contrast to controlling a nonlinear circuit through the parameter tuning of a state-space model, our approach is purely input-output. The circuit elements are controlled and interconnected to shape the current-voltage characteristics (I-V curves) of the circuit in prescribed timescales. In turn, shaping those I-V curves determines the excitability properties of the circuit. We show that this methodology enables both robust and accurate control of the circuit behavior and resembles the biophysical mechanisms of neuromodulation. As a proof of concept, we simulate a SPICE model composed of MOSFET transconductance amplifiers operating in the weak inversion regime. 
\end{abstract}


%
\IEEEpeerreviewmaketitle

\section{Introduction}
%
%
%
%
\IEEEPARstart{A}{lthough} digital technology has been evolving at an amazing pace, a large gap persists between the state-of-the-art hardware and even the simplest animal organisms at tasks that involve sensory-motor integration. Instead of relying on the standard digital electronics architecture, neuromorphic engineering aims at bridging this gap by emulating the structure of biological neurons, so that neuromorphic circuits could potentially undertake information processing in a fundamentally more efficient way \cite{mead_neuromorphic_1990,boahen_neuromorphs_2017}.

Biological neural networks possess an astonishing level of control capabilities spanning from whole brain regions controlling phenomena like attention and cognition, to ion channels that control the spiking behavior of single neurons. In particular, neuromodulators are able to modify the collective conductance of ion channels in a neuron and thus shape the neural spikes in a precise manner \cite{marder_neuromodulation_2012}. These mechanisms can lead to qualitatively different spiking regimes such as burst firing that can have a distinct function in sensory information processing \cite{marsat_behavioral_2006,fujita_spatiotemporal_2007,beurrier_subthalamic_1999,sherman_tonic_2001,krahe_burst_2004}. In particular, the transition between regular spiking and bursting oscillations is an essential mechanism of capturing the sensing scale of certain sensory systems \cite{krahe_burst_2004}. Understanding and designing simple circuits that are able to reproduce these precise controlling mechanisms could lead to novel information processing paradigms.

One of the main design choices that we face when approaching the development of neuromorphic hardware is finding the right level of abstraction of the neural behavior \cite{herz_modeling_2006}. Previous approaches have focused on replicating in silico the differential equations of neuronal conductance-based models \cite{hodgkin_quantitative_1952,farquhar_bio-physically_2005,simoni_multiconductance_2004,mahowald_silicon_1991,yu_analog_2010,hynna_thermodynamically_2007}, or reduced models such as FitzHugh-Nagumo \cite{fitzhugh_impulses_1961,keener_analog_1983,linares-barranco_cmos_1991} and integrate-and-fire \cite{izhikevich_simple_2003,indiveri_vlsi_2006,wijekoon_compact_2008,arthur_silicon-neuron_2011}. The trade-off is often between the level of biophysical details accounted for and the circuit complexity \cite{indiveri_neuromorphic_2011}. Simple abstract models have been proposed that can reproduce many distinct spiking waveforms \cite{izhikevich_simple_2003}, but they require fine-tuning of the parameters and lack the robust and smooth neuromodulation capabilities of physiological neurons \cite{pottelbergh_robust_2018}.

The present paper departs from earlier approaches by rooting the design and analysis in an input-output rather than state-space model of the circuit. Preliminary results have been reported in \cite{ribar_bursting_2017}. We assume the specific circuit architecture common to all voltage-gated conductance-based models of neurophysiology: the excitable membrane is modeled as a passive RC circuit in parallel with (possibly many) circuit elements, each of which controls the circuit conductance in a specific voltage and dynamic range. This architecture is also common to recently introduced low-dimensional model of bursting \cite{franci_organizing_2012,franci_modeling_2014}. By separating the circuit elements in distinct timescales, we propose that shaping the circuit's I-V characteristics in those distinct timescales is modular and sufficient to control the excitability properties of the circuit. The curve shaping methodology maps with surprising ease to the dynamical behaviors of the circuit, and allows us to generalize the intuitive spike-generation mechanisms of the FitzHugh-Nagumo circuit to the more complex neuronal behaviors such as bursting, while also enabling easy modulation between distinct behaviors.

Although the proposed methodology is rooted in the rigorous mathematical analysis of a low-dimensional bursting model \cite{franci_organizing_2012,franci_modeling_2014}, the key contribution of this paper is to present a qualitative approach to tuning the neuronal behaviors which is purely input-output and entirely by-passes the state-space realization of the circuit for its design and analysis. This is in contrast with recent efforts in designing spiking and bursting feedback circuits from simplified state-space models \cite{franci_realization_2014, castanos_implementing_2017}. Here we directly formulate the task of controlling a given excitable behavior as an I-V curve synthesis problem, independent of the circuit implementation specifics. Most importantly, our circuits are inherently neuromodulable as the control of each current element directly maps to the modulation of a maximal conductance parameter in biophysical conductance-based models \cite{drion_dynamic_2015,goldman_global_2001,oleary_cell_2014,swensen_robustness_2005}. This is regarded as a key step towards neuromorphic circuits with neuromodulation capabilities. As a proof of concept, we also discuss how the circuit structure can be implemented in hardware and include a SPICE simulation of the proposed circuit.

Our paper is organized as follows: We start by describing the general structure of the neural circuit, relating it to standard examples of excitable circuits such as the FitzHugh-Nagumo model, as well as more recent bursting models. We then define the notion of I-V curves in separate timescales, and describe a simple yet general circuit architecture that allows us to shape the I-V curves through parallel interconnections of basic elements with localized conductance and first-order dynamics. In the following sections, we show how, similarly to the classic FitzHugh-Nagumo circuit, a tunable spiking neuron is realized as an interconnection of the passive membrane, a fast negative conductance element, and a slow positive conductance element. We further show how the interconnection of additional slow negative conductance and ultra-slow positive conductance elements leads to a tunable bursting circuit, mirroring the ionic conductance structure of the bursting neurons \cite{rinzel_dissection_1987,drion_neuronal_2015}. We also compare the bursting mechanism we present with alternative mechanisms studied in neurodynamical models \cite{rinzel_analysis_1989}. We show why such mechanisms do not allow for robust control due to their fragile parameter choice requirements. Finally, we propose a circuit implementation of the localized conductance element, and present a feasibility study through a SPICE simulation of the circuit using the TSMC \SI{0.35}{\micro\meter} process parameters.

\section{I-V curve shaping by interconnection}
\label{localized_conductance}

\begin{figure}[!t]
\centering
\includegraphics[width=1\linewidth]{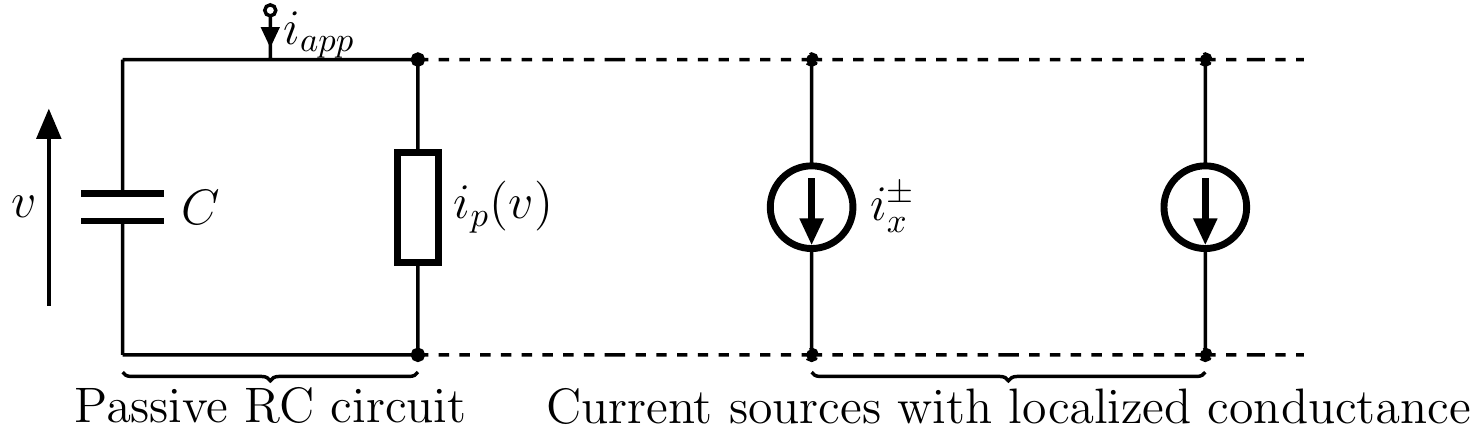}
\caption{The neural circuit: A passive RC circuit is interconnected with localized conductance current source elements that model the action of the ionic currents.}
\label{fig:neural_circuit}
\end{figure}

We base our methodology on the general neural circuit architecture shown in Fig. \ref{fig:neural_circuit}. This architecture mirrors the biophysical modeling principles pioneered by Hodgkin and Huxley in their seminal work \cite{hodgkin_quantitative_1952}: an excitable membrane is modeled as the parallel interconnection of a passive RC circuit with several voltage-gated ionic currents (sodium, potassium, calcium, etc.). Here, the passive membrane properties are represented with a membrane capacitor $C$ and a purely resistive element $i_p(v)$, so that its I-V characteristic satisfies:
\begin{equation}
\frac{di_p(v)}{dv} \geq 0, \forall v
\end{equation}

We assume that each voltage-controlled current source obeys the elementary model:
\begin{subequations}
\label{eq:current_element}
\begin{align}
i_x^{\pm} &= f_x^{\pm} (v_x) \\
T_x \dot{v}_x &= v - v_x
\end{align}
\end{subequations}
so that the output current $i_x^{\pm}$ has a monotonic dependence on the filtered voltage $v_x$ through the function $f_x^{\pm} (v)$ that satisfies
\begin{equation}
\frac{f_x^{+} (v)}{dv} \geq 0, \forall v
\end{equation}
for a positive conductance element, or
\begin{equation}
\frac{f_x^{-} (v)}{dv} \leq 0, \forall v
\end{equation}
for a negative conductance element. The time constant $T_x$ defines the timescale of the current.

The behavior of the circuit is then governed simply by Kirchhoff's law:
\begin{equation}
\label{eq:kirchhoff}
C \dot{v} = -i_p(v) -\sum i_x^{\pm}  +i_{app},
\end{equation}
and we can define an approximate time constant of the voltage equation:
\begin{equation}
\label{eq:membrane_time_constant}
C / T_v = \frac{di_p(v)}{dv}\bigg|_{v=v_e}
\end{equation}
with the derivative taken around the equilibrium point of the system $v = v_{e}$. This allows us to consider all time constants relative to the membrane dynamics:
\begin{equation}
\tau_x = T_x / T_v
\end{equation}

In order to describe both bursting and spiking behaviors, it is sufficient to include elements that act on three separate timescales: fast ($\tau_f$), slow ($\tau_s$) and ultra-slow ($\tau_{us}$), for which we assume:
\begin{equation}
\max(\tau_v , \tau_f) \ll \tau_s \ll \tau_{us}
\end{equation}
The dynamics of the fast element relative to the membrane dynamics can be arbitrary, and in particular, the fast dynamics can be taken as instantaneous. We use an instantaneous fast element throughout the simulations in the paper, corresponding to $\tau_f = 0$.

For modeling purposes, we will consider a dimensionless representation of this circuit architecture. We will denote the dimensionless quantities with capital letters, so that equations \eqref{eq:current_element} and \eqref{eq:kirchhoff} become:
\begin{subequations}
\label{eq:dimensionless}
\begin{align}
\frac{dV}{d\tau} &= -I_p(V) -\sum I_x^{\pm} + I_{app} \\
I_x^{\pm} &= F_x^{\pm} (V_x) \\
\tau_x \frac{dV_x}{d\tau} &= V - V_x,
\end{align}
\end{subequations}
with $\tau$ being the dimensionless time.

The classic FitzHugh-Nagumo model of excitability assumes this form with the elements:
\begin{subequations}
\begin{align}
I_p(V) &= V^3 / 3 \\
F_f^-(V) &= -V \\
F_s^+(V) &= kV,
\end{align}
\end{subequations}
with $\tau_f = 0$ and $k > 1$. The recent model \cite{franci_modeling_2014} generalizes FitzHugh-Nagumo circuit to allow for a modulation between bursting and spiking behaviors. It assumes the form:
\begin{subequations}
\begin{align}
I_p(V) &= V^3 / 3 \\
F_f^-(V) &= -V \\
F_s^-(V) + F_s^+(V) &= (V + V^*)^2 \\
F_{us}^+(V) &= V,
\end{align}
\end{subequations}
The bursting behavior relies on a non-monotonic slow current and an additional ultra-slow current. The parameter $V^*$ controls if the model is spiking or bursting.

Here we will consider those particular circuits as specific interconnections of standardized elements with I-V characteristic
\begin{equation}
\label{eq:localized_conductance}
F_x^{\pm}(V) = \pm \alpha_x^{\pm} \tanh (V_x - \delta_x^{\pm}).
\end{equation}
Such characteristics retain a fundamental property of biophysical circuits: the conductance has a \textit{localized} activation range, as well as a well-defined timescale. The local range is specified by the linear range of the sigmoid, whereas the timescale is specified by the time constant $\tau_x$. The parameter $\alpha_x^{\pm} > 0$ controls the gain of the conductance, and the parameter $\delta_x^{\pm}$ determines where in the voltage range the element is active. In addition, we have the passive element taking form of a resistor, so that:
\begin{equation}
\label{eq:passive}
I_p(V) = V
\end{equation}

We view the role of localized conductance elements as \textit{shaping} the I-V characteristics of the circuit in distinct timescales. Accordingly, we will consider the I-V characteristics of the circuit in the respective \textbf{fast}, \textbf{slow}, and \textbf{ultra-slow} timescales. Those curves will be denoted by:
\begin{equation}
\mathcal{I}_x = I_p(V) + \sum\limits_{\tau_y \le \tau_x} F_y^{\pm}(V),
\end{equation}
so that $\mathcal{I}_x$ represents the summation of all the I-V curves of elements acting on the timescale $\tau_x$, or faster.

The basic rationale of our design will be the following: we will use \textit{negative} conductance elements to create local ranges of negative conductance in a given timescale and \textit{positive} conductance elements to restore the positive conductance in slower timescales. The circuit behavior will be determined by shaping the local ranges of negative conductance in the right voltage ranges and timescales. Our methodology is obviously qualitative in nature: it does not depend on specific circuit or mathematical realizations but only on shaping the monotonicity properties of the I-V curves in distinct timescales.

\section{Shaping an excitable circuit}
\label{sec:excitable}

Ever since the early works of Van der Pol\cite{van_der_pol_lxxxviii._1926}, Hodgkin and Huxley \cite{hodgkin_quantitative_1952} and FitzHugh and Nagumo \cite{fitzhugh_impulses_1961,nagumo_active_1962}, the property of excitability of a circuit has been rooted in a region of \textit{negative} conductance in a specific voltage range \cite{sepulchre_excitable_2018}. As a first step, we briefly revisit this construction with our I-V shaping technique.

\begin{figure}[!t]
\centering
\includegraphics[width=0.8\linewidth]{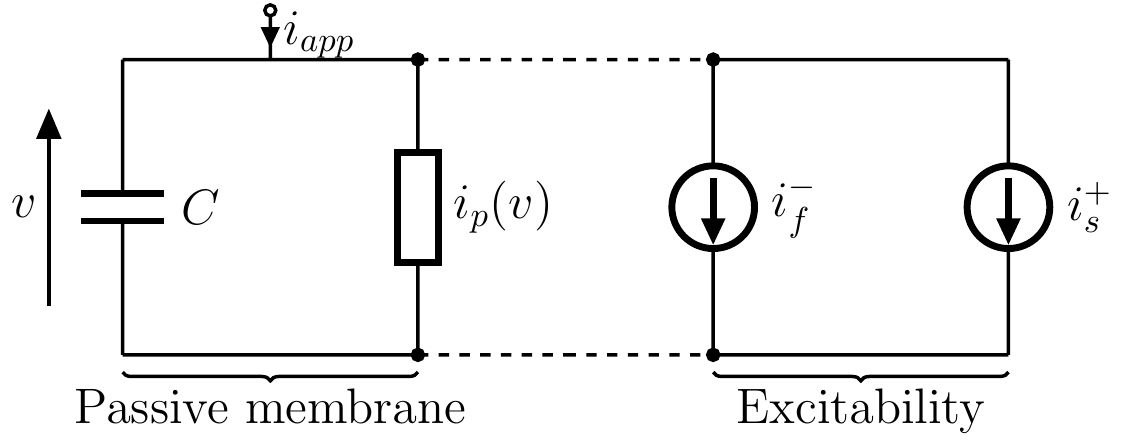}
\caption{Synthesis of an excitable circuit as the parallel interconnection of a passive membrane with a fast negative conductance element ($I_f^-$), balanced by a slow positive conductance element ($I_s^+$).}
\label{fig:excitable_circuit}
\end{figure}

\begin{figure}[!t]
\centering
\includegraphics[width=1\linewidth]{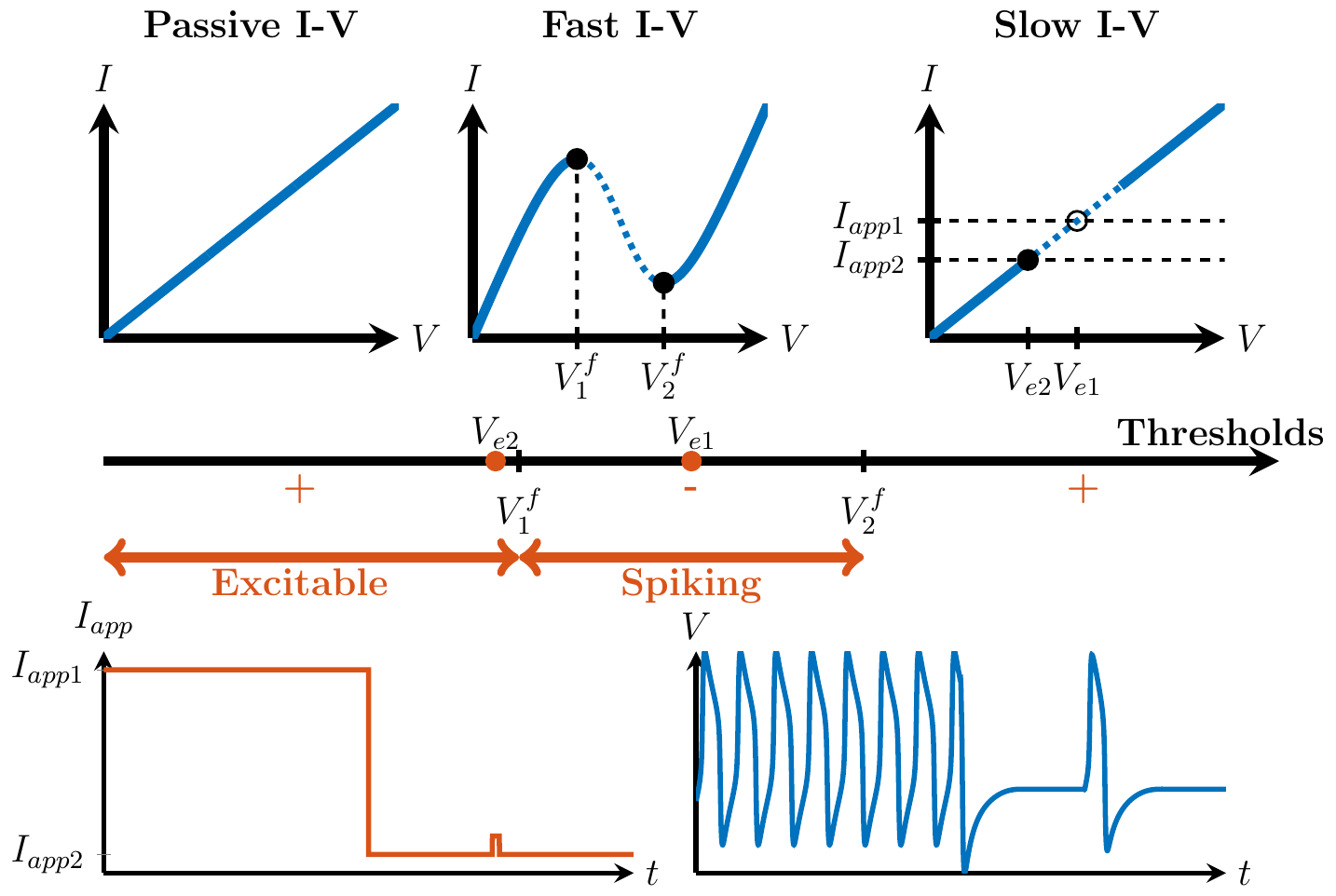}
\caption{Properties of an excitable circuit. Top: Passive, fast and slow I-V curves of an excitable circuit. The fast curve is ``N-shaped''. The slow curve is monotone and its intersection with the line $I = I_{app}$ determines the system's equilibrium ($V_e$). Middle: If the equilibrium voltage lies in the negative conductance region of the fast I-V curve ($V_{e1}$) the system is spiking, or is excitable otherwise ($V_{e2}$). Voltage regions are indicated with the sign of the slope of the fast I-V curve. Bottom: Transition between the spiking and excitable regimes through the applied current.}
\label{fig:excitable_currents}
\end{figure}

The excitable circuit in Fig. \ref{fig:excitable_circuit} uses two localized conductance elements: a fast negative conductance element $I_f^-$ and a slow positive conductance element $I_s^+$. The role of $I_f^-$ is to create a range ($V_1^f$,$V_2^f$) in the total fast I-V curve. The role of $I_s^+$ is to restore a positive conductance characteristic in the total slow I-V curve. The I-V curve shaping is thus determined by the following two conditions: 
\begin{equation}
\frac{d\mathcal{I}_f}{dV} <0, V \in (V_1^f, V_2^f)
\end{equation}

\begin{equation}
\frac{d\mathcal{I}_s}{dV} > 0, \forall V
\end{equation}
which correspond to the graphs illustrated in Fig. \ref{fig:excitable_currents}: the passive and slow I-V curves are monotone, whereas the fast I-V curve has the characteristic ``N-shape'' of a negative conductance circuit.

%

Provided that the timescale separation is sufficient, the resulting dynamical behavior of the circuit has the following properties:

\begin{itemize}
\item The circuit has a monotone equilibrium characteristic given by the slow I-V curve. A unique equilibrium ($I_{app}$,$V_e$) exists for every value of $I_{app}$. The equilibrium voltage is stable except in a finite range included in the interval ($V_1^f$,$V_2^f$).
\item The circuit has a stable spiking behavior in a finite range of constant applied current. The spiking behavior is characterized by a stable limit cycle oscillation with sharp upstrokes and downstrokes between a ``low'' and a ``high'' voltage range.
\item For equilibrium voltages close to the negative conductance range ($V_1^f$,$V_2^f$), the circuit is excitable: the steady-state behavior is a stable equilibrium but small current pulses can trigger ``spikes'', i.e. a transient manifestation of the oscillatory behavior.
\end{itemize}

\begin{figure}[!t]
\centering
\includegraphics[width=0.8\linewidth]{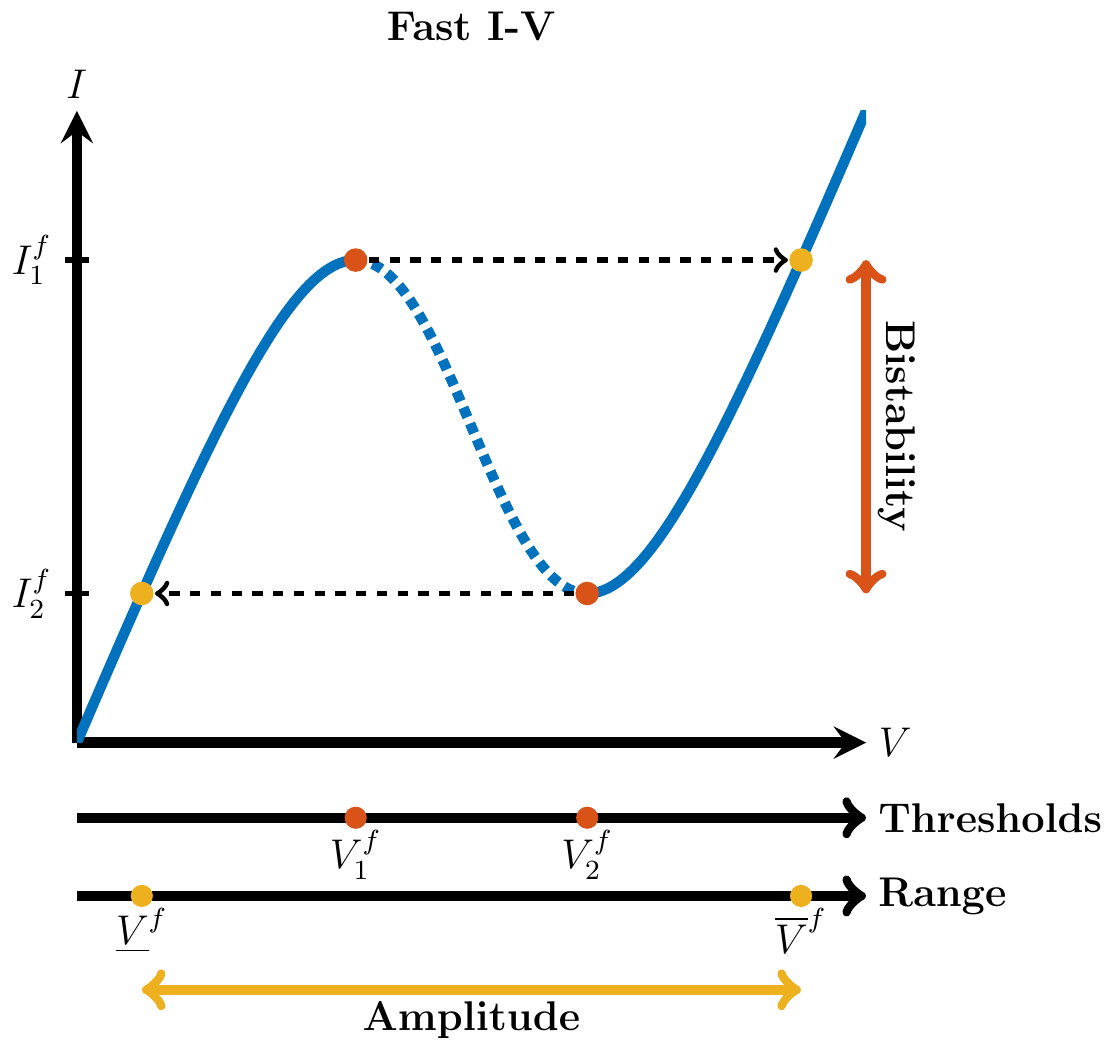}
\caption{The ``N-shaped'' fast I-V curve. The threshold voltages $V_1^f$ and $V_2^f$ define the bistable region, so that by increasing the current above $I_1^f$ or decreasing it below $I_2^f$ makes the system jump to the opposite branch of the curve, defined by voltages $\overline{V}^f$ and $\underline{V}^f$ respectively. Controlling the location of these points represents the essential modulation mechanism of the circuit.}
\label{fig:excitable_bistability}
\end{figure}

The three properties above determine an excitable behavior \cite{sepulchre_excitable_2018}. They primarily owe to the fast-slow decomposition of the circuit. In the fast timescale, the negative conductance characteristic makes the circuit bistable and hysteretic: in the range of currents ($I_1^f$,$I_2^f$), two stable voltage points coexist, so that the behavior can easily switch between a low voltage state in the range ($\underline{V}^f$,$V_1^f$) and a high voltage state in the range ($V_2^f$,$\overline{V}^f$) (Fig. \ref{fig:excitable_bistability}). In the slow timescale, the positive conductance characteristic makes the circuit monostable, resulting in either a stable equilibrium or a stable spiking behavior.

This analysis is consistent with the biophysics of excitable neurons: sodium channel activation is fast and acts as a negative conductance close to the resting potential, whereas potassium channel activation is slow and acts as a positive conductance.

\section{Neuromodulating an excitable circuit}

\begin{figure}[!t]
\centering
\includegraphics[width=1\linewidth]{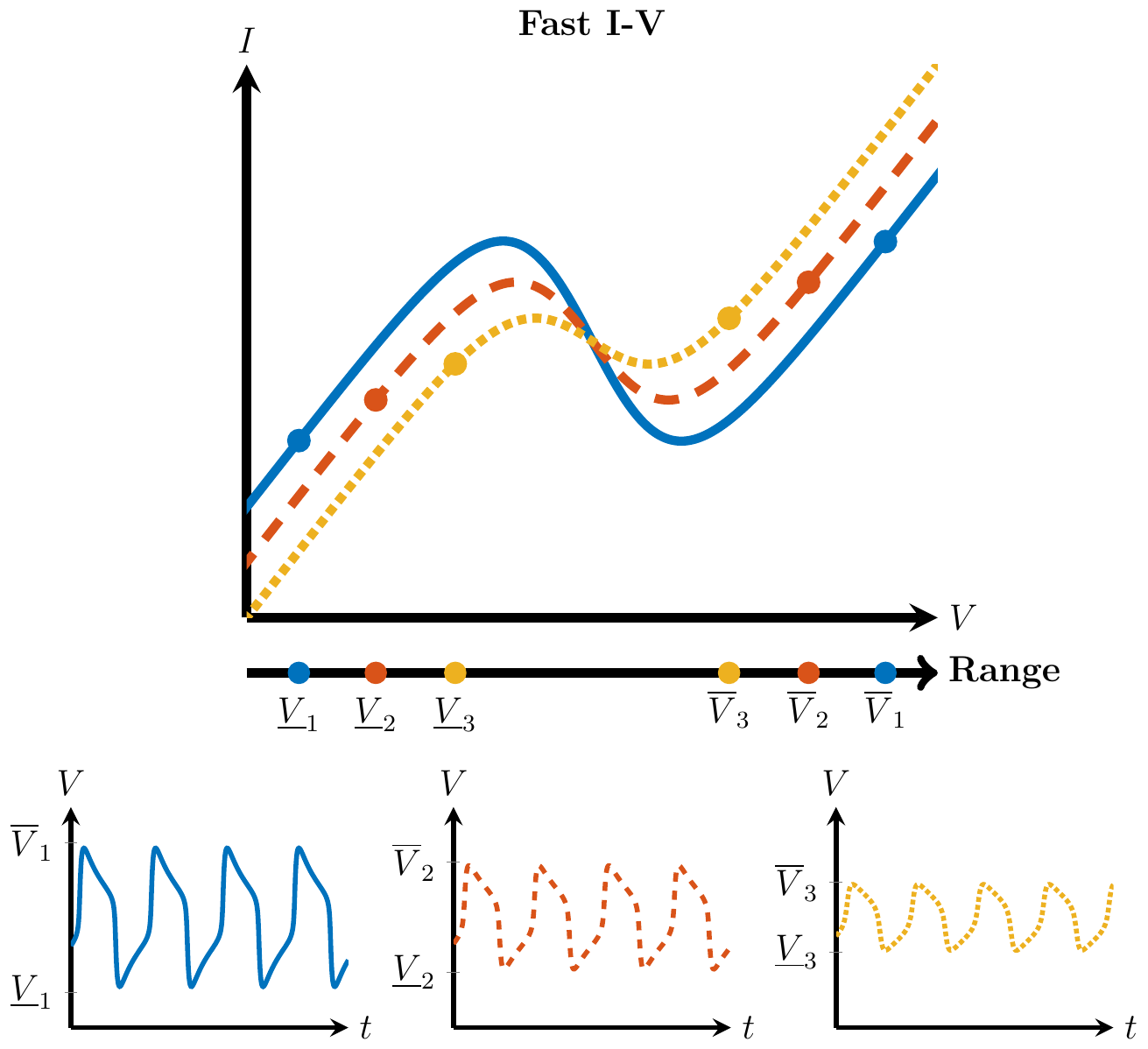}
\caption{Amplitude of the spikes is determined by the fast I-V curve. The jumps between the low and the high voltage happen at the local maximum and minimum of the curve. Increasing the gain of the fast negative conductance element ($\alpha_f^-$) widens the negative conductance region and increases the amplitude of the spikes. In order to keep the frequency of the oscillations approximately constant, the gain of the slow positive conductance element is kept the same as the gain of the slow negative conductance in all simulations ($\alpha_s^+ = \alpha_f^-$).}
\label{fig:excitable_amp}
\end{figure}

\begin{figure}[!t]
\centering
\includegraphics[width=1\linewidth]{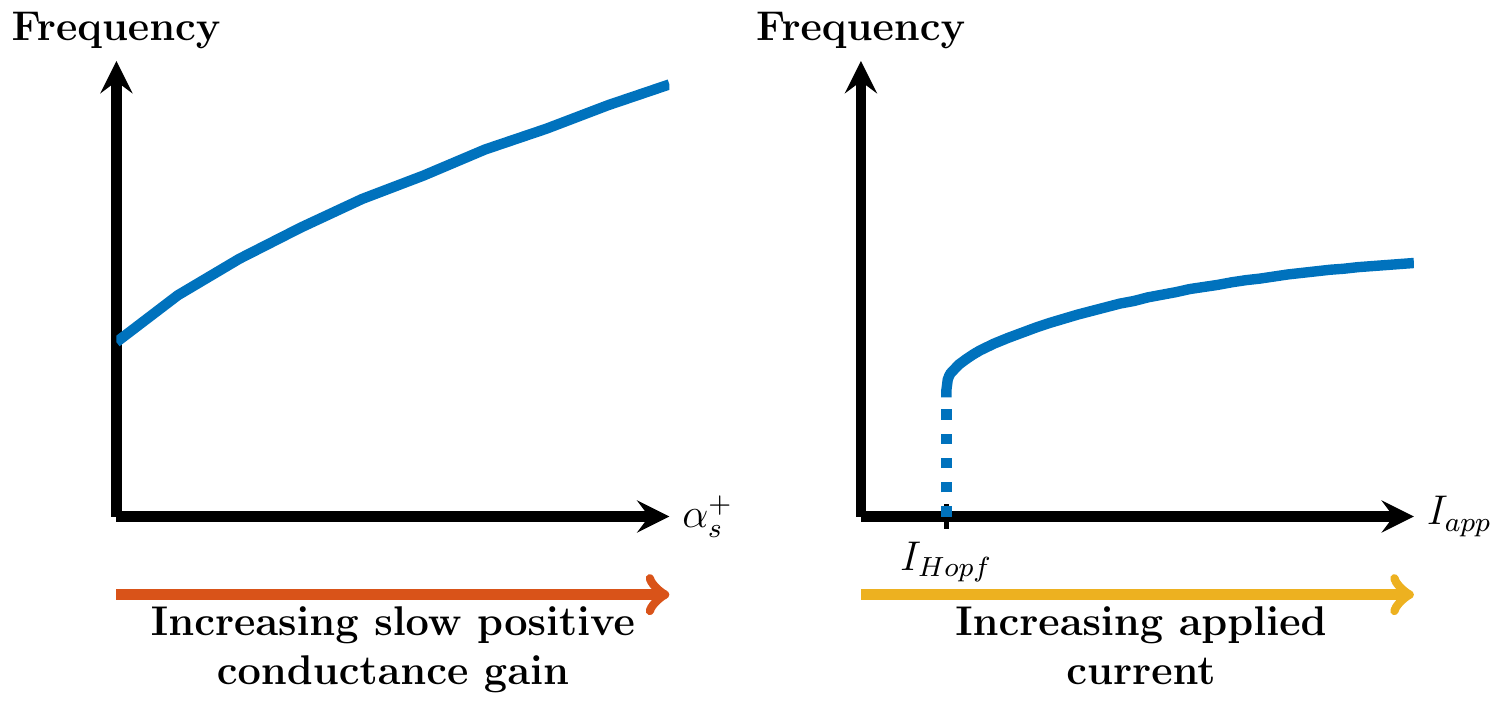}
\caption{Tuning the frequency of spiking. Left: Frequency of the oscillation can be controlled by varying the slow positive conductance gain ($\alpha_s^+$). This controls the ``up'' and ``down'' time intervals of the spike, which approximately determines the period of oscillation due to the fast nature of the jumps. Right: For fixed $\alpha_s^+$, increasing the applied current increases the frequency, but there is a discontinuous jump at $I_{Hopf}$ due to the oscillations emerging through a Hopf bifurcation.}
\label{fig:excitable_frequency}
\end{figure}

Provided a sufficient timescale separation between ``fast'' and ``slow'', the qualitative behaviors of our excitable circuit are solely determined by the I-V curve shaping. Classical dynamical systems tools (see e.g. \cite{izhikevich_dynamical_2007}) show that the unstable voltage range is delineated by two Hopf bifurcations and that the unstable range converges to ($V_1^f$,$V_2^f$) as the ratio $\max(\tau_v,\tau_f) / \tau_s$ approaches zero. These asymptotic properties are very convenient to tune the excitable circuit from its fast and slow I-V curves:
\begin{itemize}
\item The amplitude range of the spiking behavior is determined by the hysteresis of the fast I-V curve. With a localized conductance like \eqref{eq:localized_conductance}, the hysteresis is centered around $\delta_f^-$ and its range is modulated by the control parameter $\alpha_f^-$ (Fig. \ref{fig:excitable_amp}).
\item The spiking frequency is determined by the time spent in the low and high voltage range ($\underline{V}^f$,$V_1^f$) and ($V_2^f$,$\overline{V}^f$), respectively. For a fixed negative conductance element, this time is primarily modulated by the control parameter $\alpha_s^+$ (Fig. \ref{fig:excitable_frequency}, left). The frequency can also be modulated by the applied current (Fig. \ref{fig:excitable_frequency}, right). The neuron is of Type II in the terminology of \cite{izhikevich_dynamical_2007} because of a nonzero minimal spiking frequency at the Hopf bifurcation.
\end{itemize}

\section{Shaping a bursting circuit}

\begin{figure}[!t]
\centering
\includegraphics[width=1\linewidth]{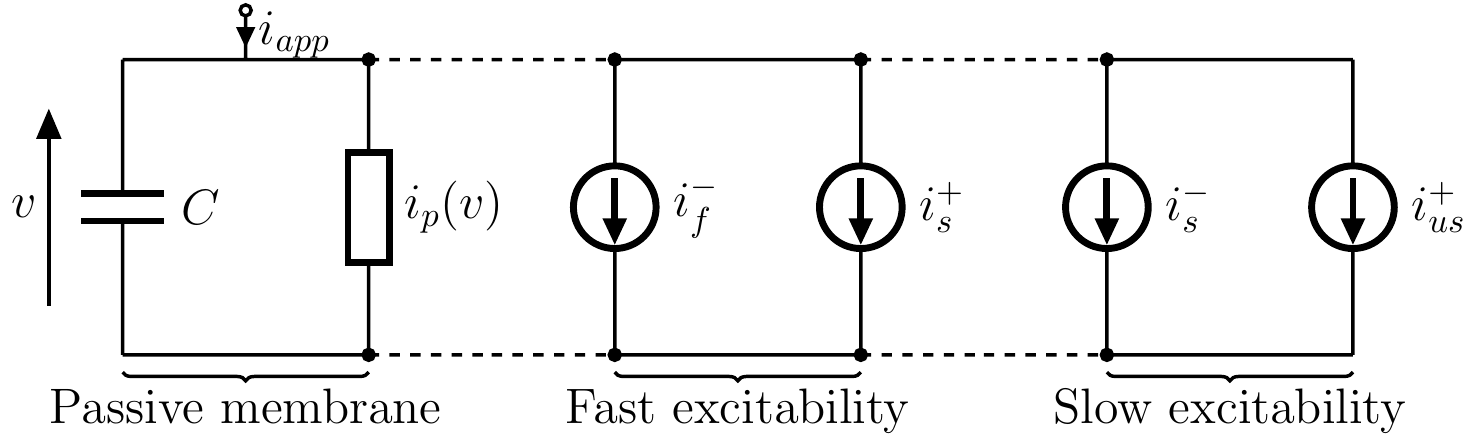}
\caption{Synthesis of a bursting circuit as the parallel interconnection of a passive membrane with both fast ($I_f^-$) and slow ($I_s^-$) negative conductance elements, respectively balanced by slow ($I_s^+$) and ultra-slow ($I_{us}^+$) positive conductance elements.}
\label{fig:bursting_circuit}
\end{figure}

Our bursting circuit in Fig. \ref{fig:bursting_circuit} closely mimics the architecture of the spiking circuit in the previous section. We view bursting as shaped by two, rather than one, ranges of negative conductance: the first one in the fast timescale, created by $I_f^-$, and a second one in the slow timescale, created by $I_s^-$. The first negative conductance is balanced by a positive conductance $I_s^+$ in the slow timescale, whereas the slow negative conductance is balanced by a positive conductance $I_{us}^+$ in the ultra-slow timescale.


\begin{figure}[!t]
\centering
\includegraphics[width=1\linewidth]{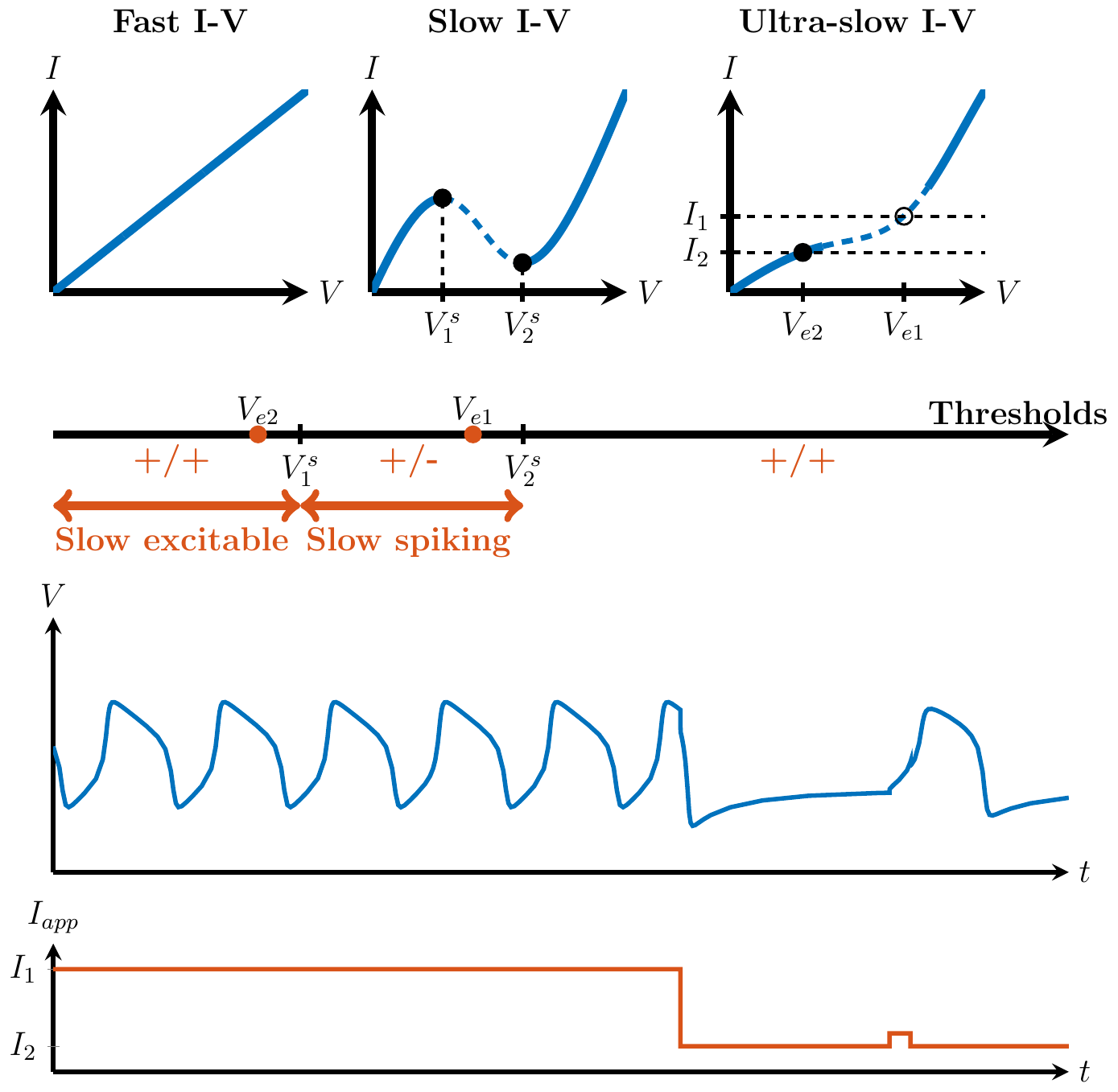}
\caption{Slow excitable circuit. Without the fast excitability elements, fast I-V curve is monotonic, slow I-V curve is ``N-shaped'', and the ultra-slow I-V curve is monotonic, so that the system is slow excitable, similarly to Fig. \ref{fig:excitable_currents}. The voltage regions are now indicated with two signs, so that the first sign corresponds to the sign of the slope of the fast I-V curve, and the second sign corresponds to the sign of the slope of the slow I-V curve.}
\label{fig:slow_spiking}
\end{figure}

Fig. \ref{fig:slow_spiking} illustrates the design of the slow excitable circuit (in the absence of $I_f^-$ and $I_s^+$) exactly as in the previous section. 

\begin{figure}[!t]
\centering
\includegraphics[width=1\linewidth]{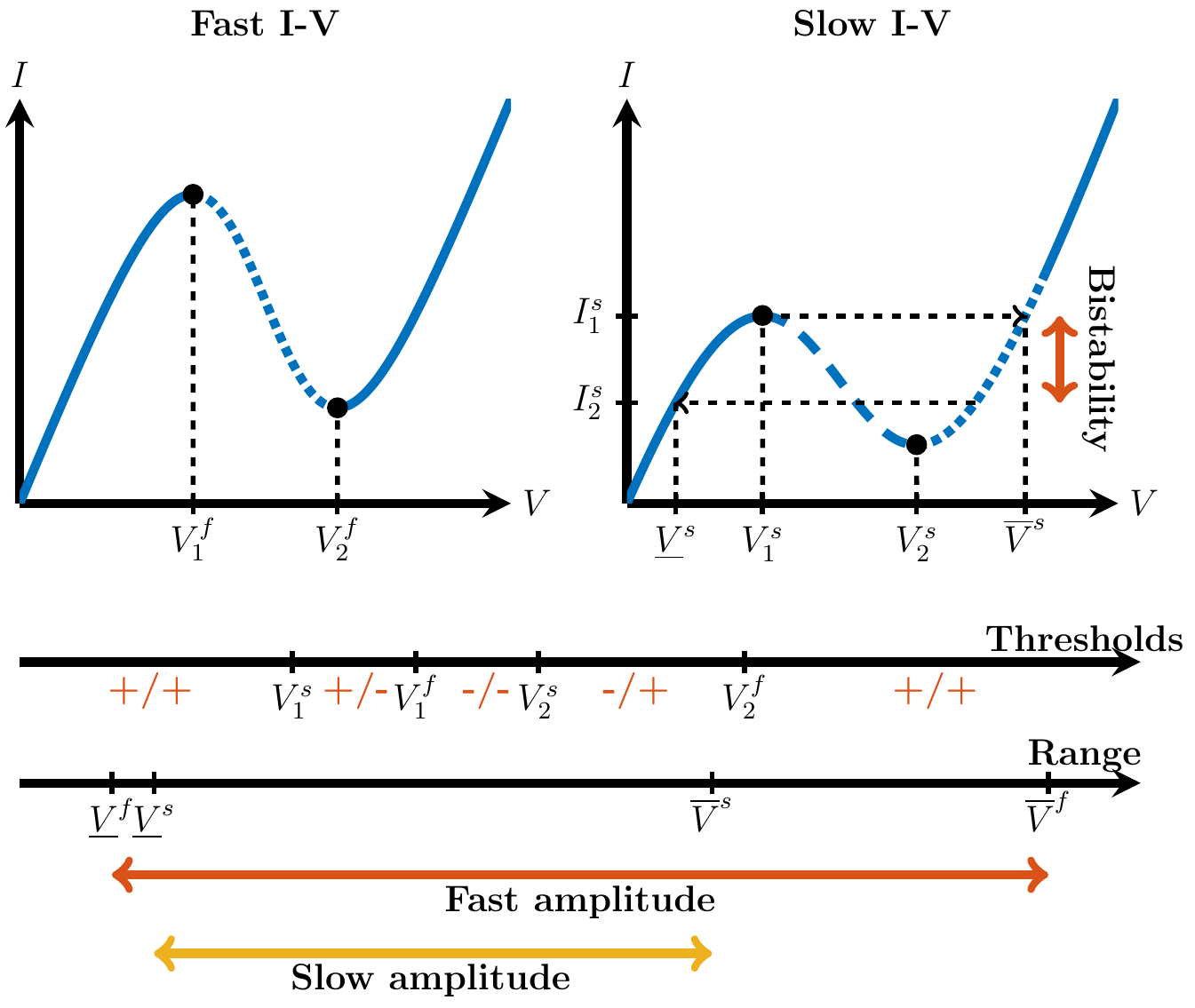}
\caption{Slow bistability between the rest state and the spiking state. The I-V curves (top) show a system with a double hysteresis: both the fast and the slow I-V curves are ``N-shaped''. By having the ``up'' state of the slow curve correspond to the unstable region of the fast system, the system experiences rest-spike bistability, given that the slow threshold is at lower voltage than the fast one, i.e. $V_1^s < V_1^f$. The system now has two pairs of threshold and range voltages corresponding to the fast and the slow I-V curves, which gives the full set of modulation variables for controlling the behavior of the circuit.}
\label{fig:slow_bistability}
\end{figure}

\begin{figure}[!t]
\centering
\includegraphics[width=1\linewidth]{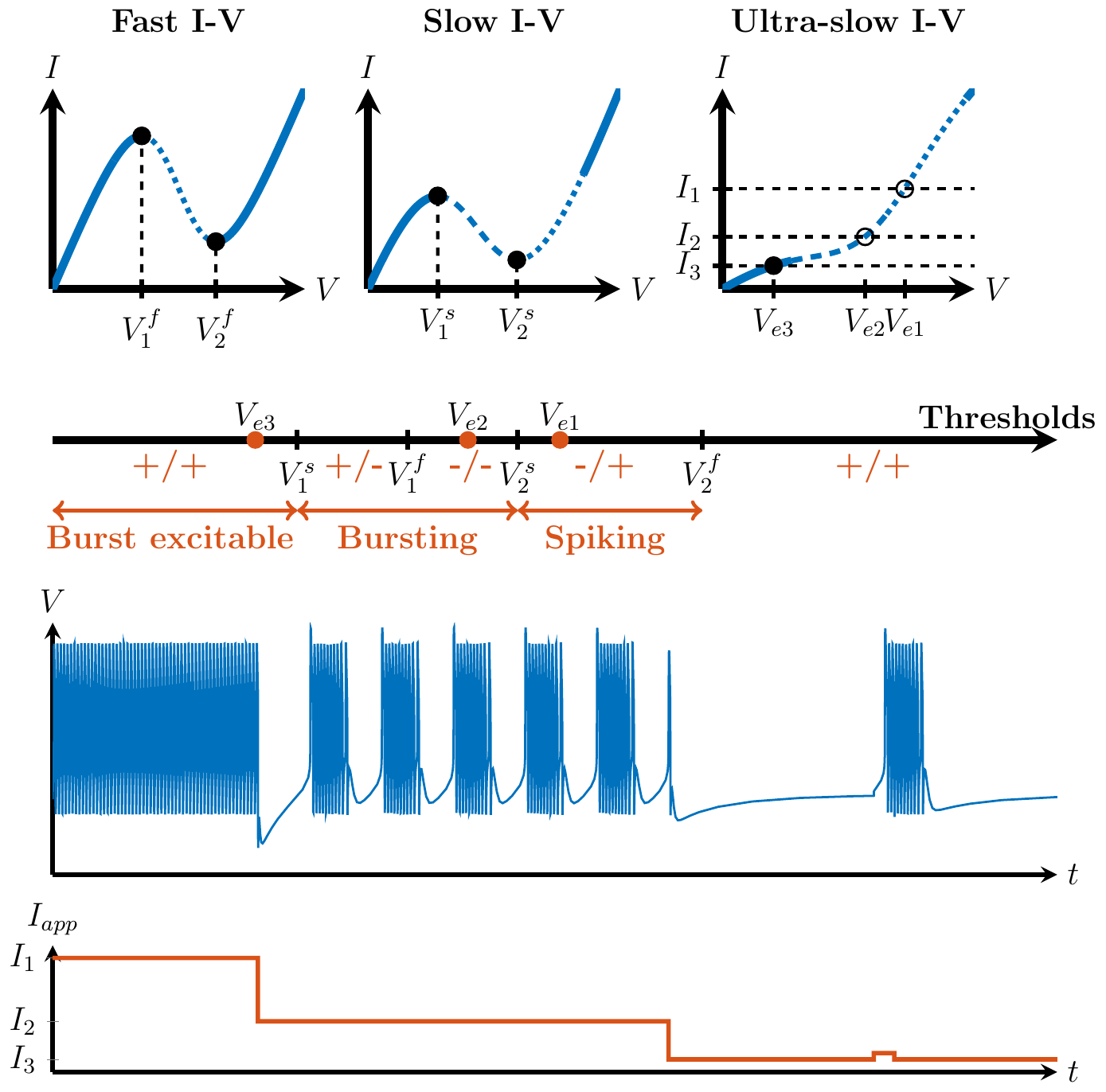}
\caption{Properties of a bursting circuit. Top: Both fast and slow I-V curves are ``N-shaped''. The ultra-slow I-V curve is monotonically increasing, and the intersection with the $I=I_{app}$ line determines the location of the system's equilibrium $V_e$. Middle: If the equilibrium voltage lies below the slow threshold, the system is burst excitable ($V_{e3}$), if the equilibrium lies in the negative conductance region of the slow I-V curve the system is bursting ($V_{e2}$), while if the system lies in the negative conductance region of the fast I-V curve above $V_2^s$ ($V_{e1}$), the system is purely fast spiking. Bottom: The transition between the three regimes through $I_{app}$.}
\label{fig:bursting}
\end{figure}

Bursting is obtained by shaping the fast and slow I-V curves as illustrated in Fig. \ref{fig:slow_bistability}. Each curve has a range of negative conductance and the two ranges overlap in such a way that
\begin{equation}
V_1^s < V_1^f < V_2^s < V_2^f.
\end{equation}
Finally, the ultra-slow positive conductance element restores monotonicity in the ultra-slow I-V curve, as illustrated in Fig. \ref{fig:bursting}. Provided that the timescale separation is sufficient, the resulting dynamical behavior has the following properties:
\begin{itemize}
\item The circuit has a monotone equilibrium characteristic given by the ultra-slow I-V curve.
\item Depending on the constant applied current, the circuit has a stable equilibrium (resting state), a stable limit cycle behavior (spiking), or a stable limit cycle characterized by an alternation of spikes and rest (bursting state).
\item Close to the bursting range of applied currents, the circuit is burst excitable. The steady-state behavior is a stable equilibrium but small current pulses can trigger individual ``bursts'', i.e. a transient manifestation of the bursting behavior.
\end{itemize}

It is remarkable that those qualitative properties are entirely determined by the shaping of I-V curves as in Fig. \ref{fig:bursting}. This owes to the three timescale decomposition of the dynamical behaviors. A detailed analysis in \cite{franci_organizing_2012,franci_modeling_2014} shows that the shaping in Fig. \ref{fig:slow_bistability} is sufficient to enforce bistability in the slow timescale between rest and spiking. The rest-spike bistable range in Fig. \ref{fig:slow_bistability} is analogous to the rest-rest bistable range in Fig. \ref{fig:excitable_bistability}. It is governed by a transcritical bifurcation at the current $I_2^s$, as studied in \cite{franci_organizing_2012,franci_modeling_2014} and derived in the Appendix \ref{appendix:a}.

The architecture of the bursting circuit is once again consistent with the biophysics of bursting neurons: the slow negative conductance is provided by the slow activation of calcium ions or the slow inactivation of potassium ions, while calcium-activated potassium channels provide the ultra-slow positive conductance.

\section{Neuromodulating a bursting circuit}

\begin{figure}[!t]
\centering
\includegraphics[width=1\linewidth]{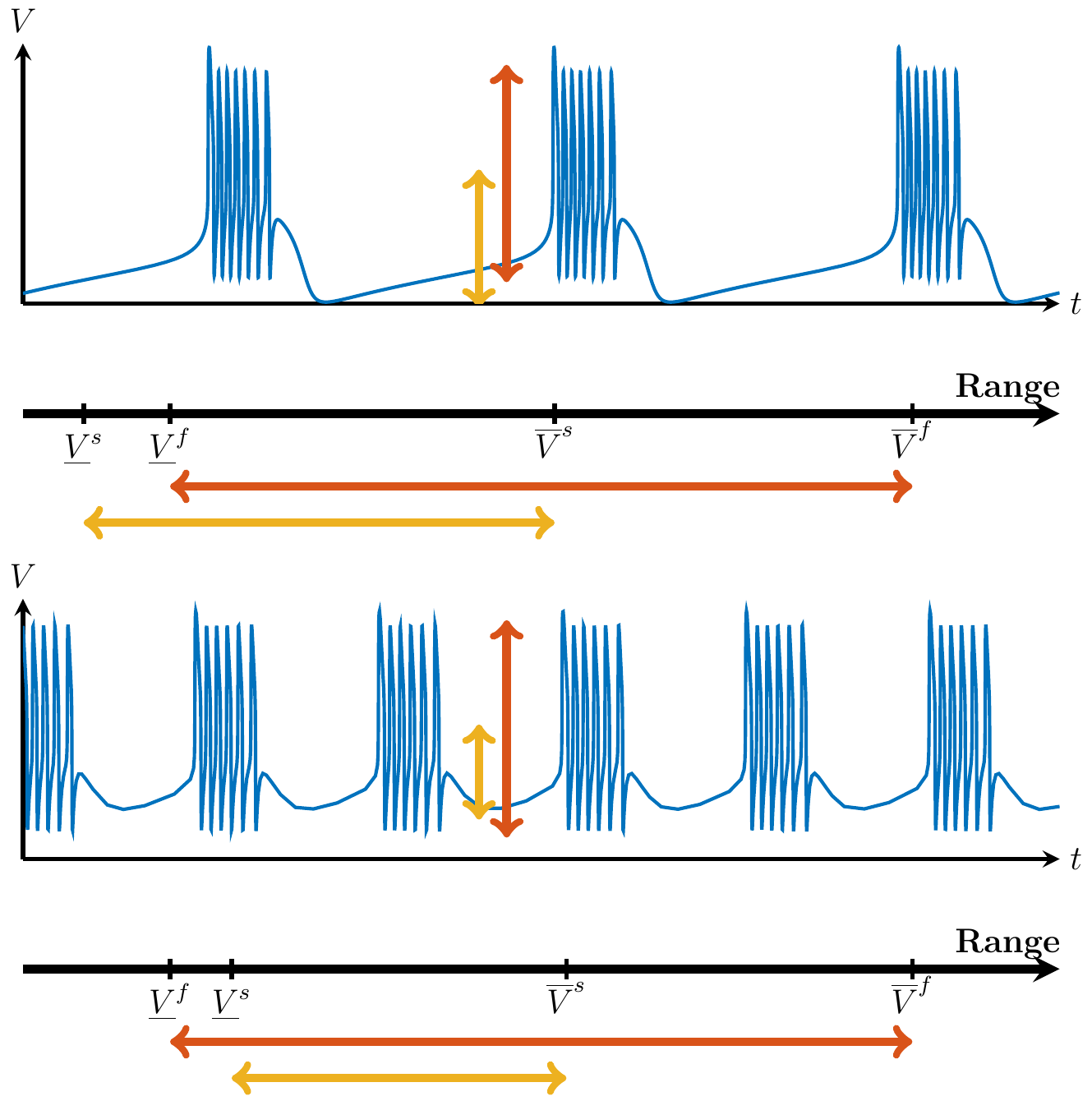}
\caption{Controlling the bursting waveform. The bursting oscillations can be designed by independently considering the fast and the slow I-V curves, and shaping both fast and slow spiking. In this way we can design plateau oscillations (top) and non-plateau oscillations (bottom), by moving the negative conductance region of the slow I-V curve relative to the the negative conductance region of the fast I-V curve (see Fig. \ref{fig:slow_bistability}).}
\label{fig:bursting_amplitude}
\end{figure}

Very much like how the fast I-V curve determined the amplitude of spiking in Fig. \ref{fig:excitable_amp}, the fast and slow I-V curves of Fig. \ref{fig:slow_bistability} determine the amplitude tuning of bursting. The gains of the negative conductance elements $I_f^-$ and $I_s^-$ can be modulated to control the amplitude properties of the bursting waveforms. For instance Fig. \ref{fig:bursting_amplitude} illustrates how moving the negative conductance regions of the fast and slow I-V curves relative to each other as well as modulating their widths, leads to a transition between plateau and non-plateau bursting.

\begin{figure}[!t]
\centering
\includegraphics[width=1\linewidth]{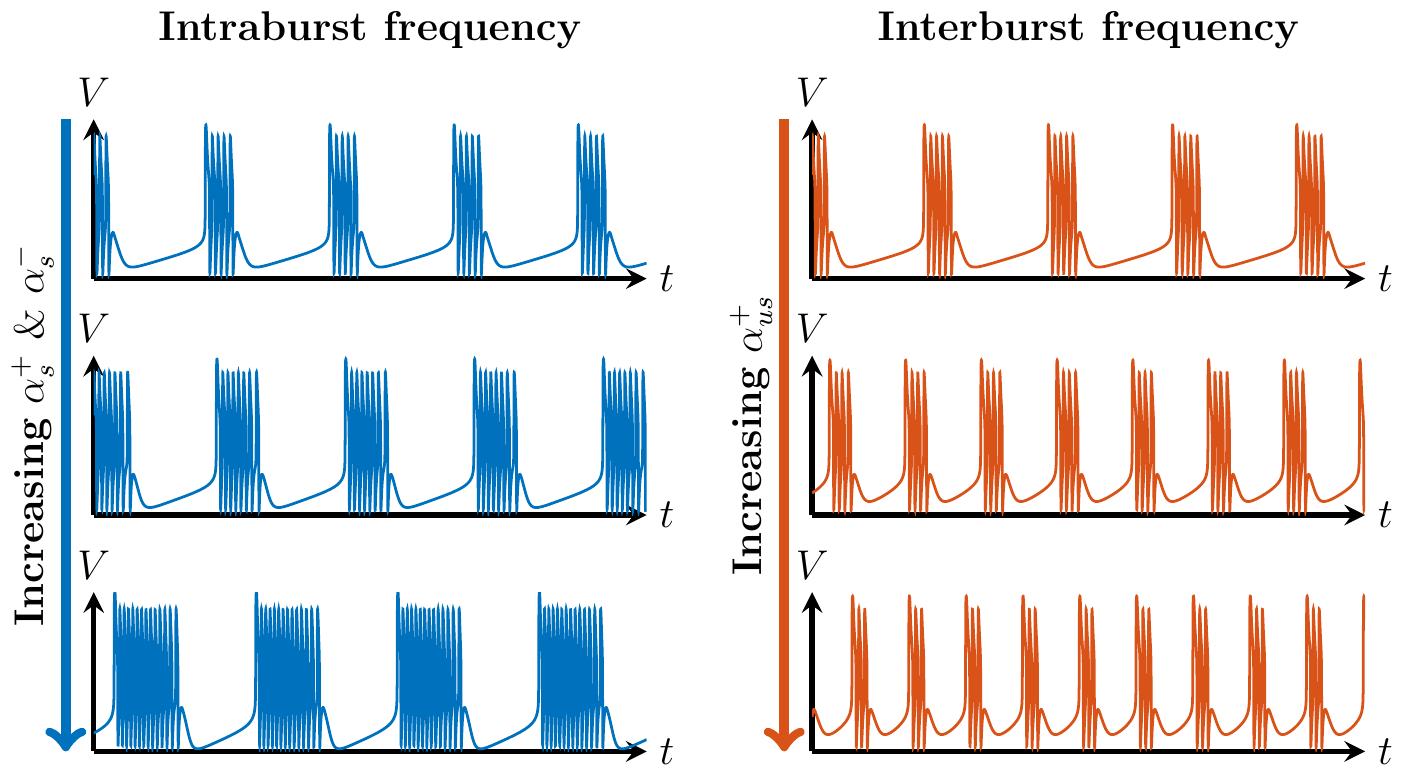}
\caption{Controlling the intraburst and interburst frequencies of the circuit. Increasing the gain of the slow positive conductance element increases the frequency of the fast spiking, thus increasing the intraburst frequency (left column); the slow negative conductance is increased by the same amount to keep the interburst period approximately constant. Increasing the gain of the ultra-slow positive conductance element increases the frequency of the slow spiking, thus increasing the interburst frequency (right column).}
\label{fig:bursting_frequency}
\end{figure}

Likewise, for fixed negative conductance elements, the gains of the positive conductance elements provide natural parameters to control intraburst and interburst frequencies of the bursting attractor (Fig. \ref{fig:bursting_frequency}).

Figs. \ref{fig:bursting_amplitude} and \ref{fig:bursting_frequency} illustrate the versatility of our approach and the relevance of tuning a bursting circuit from I-V curves rather than through an exhaustive exploration of the parameter space.

\subsection{Bursting/spiking modulation through slow I-V curve}
\label{section:bursting_spiking}

Our bursting circuit has a simple parallel architecture with four basic control parameters: two negative conductance gains, $\alpha_f^-$ and $\alpha_s^-$, and two positive conductance gains $\alpha_s^+$ and $\alpha_{us}^+$. Each negative conductance gain controls one mode of excitability : spiking, a fast excitability mode ($\alpha_f^-$) and bursting, a slow excitability mode ($\alpha_s^-$). Each positive conductance gain controls the corresponding frequency: spiking frequency ($\alpha_s^+$) and bursting frequency ($\alpha_{us}^+$).

Those control parameters are in close analogy with the maximal conductances of the four typical ionic currents of a bursting neuron: sodium and calcium currents are inward currents whose activation gating variables control the negative conductances. The activation of calcium currents is often five to ten times slower than the activation of a sodium current. Potassium and calcium-activated potassium currents are outward currents whose activation variables control the positive conductances. The activation timescale of potassium and calcium are often similar, whereas the activation of calcium-activated potassium lags behind. See \cite{franci_robust_2017} for a further analysis of the physiological conductances of a neuron.

The balance between $I_s^-$ and $I_s^+$ is particularly important in the modulation of the circuit activity between spiking and bursting. The modulation of this balance shapes the monotonicity of the slow I-V curve: a monotone shape will lead to spiking behavior whereas an ``N-shaped'' curve will lead to bursting.

Further insight into this regulation is provided by a local analysis of the I-V curves around critical points. This analysis makes contact with singularity theory, that has been the key analysis tool to analyze the modulation of bursting in \cite{franci_modeling_2014}. We briefly illustrate the value of singularity theory by studying the transition from spiking to bursting around the critical point
\begin{equation}
\label{eq:thresholds}
V = V_1^f = V_1^s,
\end{equation}
obtained by aligning the fast and slow threshold of Fig. \ref{fig:slow_bistability}. The concavity of the slow I-V curve around that point locally controls the transition from bursting (locally concave) to spiking (locally convex). The transition is determined by a change of sign in the second derivative:
\begin{equation}
\label{eq:pitch}
\frac{d^2 \mathcal{I}_s}{dV^2} = 0
\end{equation}
which, together with Eq. \eqref{eq:thresholds} determines a pitchfork bifurcation in the fast-slow model \cite{franci_modeling_2014}.

\begin{figure}[!t]
\centering
\includegraphics[width=1\linewidth]{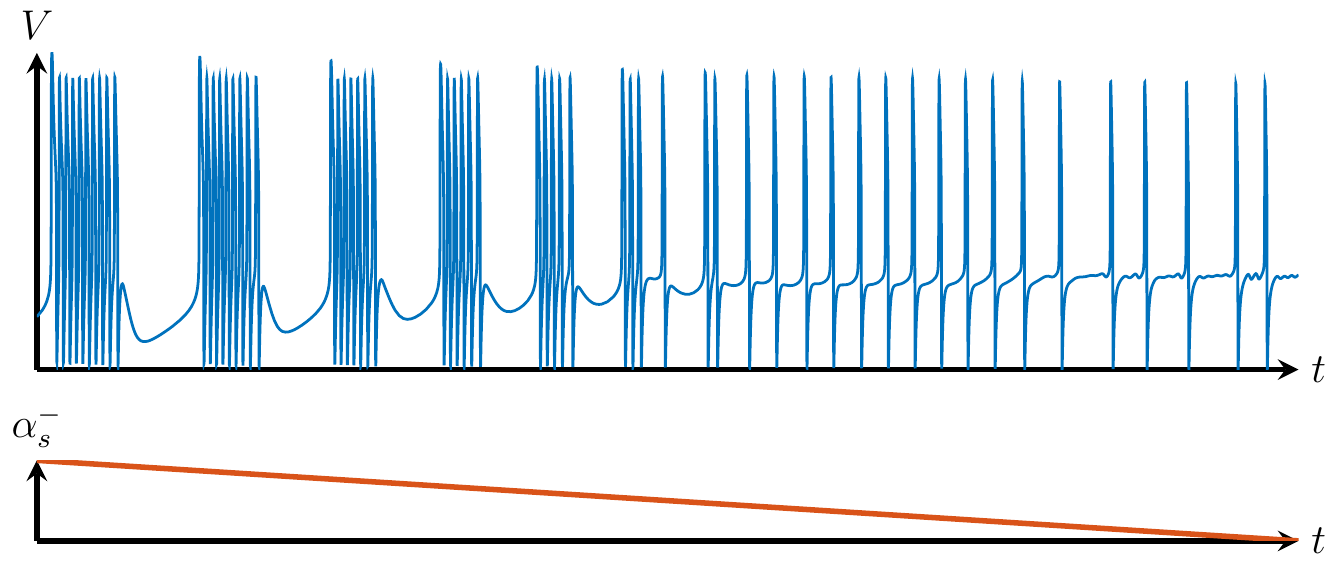}

\vspace{0.5cm}

\includegraphics[width=1\linewidth]{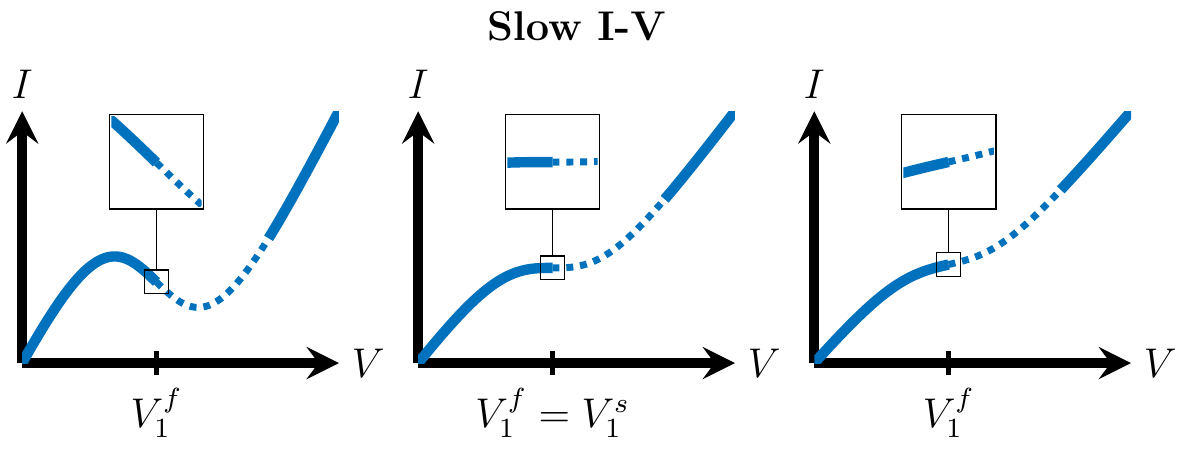}
\caption{Controlling the oscillation mode. Top: Transition between bursting and regular spiking modes by changing the gain of the slow negative conductance element. Bottom: The transition can be traced locally around the fast threshold $V_1^f$ through the circuit's slow I-V curve (bottom). Starting from a balanced condition (middle), increasing the gain makes the slope locally negative and creates slow bistability (left), while decreasing the gain makes the slow curve monotonic (right). Decreasing the size of the bistable region continuously decreases the number of spikes per burst, changing the behavior into regular spiking when bistability is lost.}
\label{fig:pitch}
\end{figure}

Fig. \ref{fig:pitch} illustrates a smooth transition from bursting to tonic spiking around that point. The transition is governed by the modulation of the sole parameter $\alpha_s^-$. Such a transition is in close analogy with the modulation of calcium currents in the physiologically significant transition from tonic spiking to bursting, see e.g. \cite{mccormick_neurotransmitter_1992}.

\begin{figure}[!t]
\centering
\includegraphics[width=0.49\linewidth]{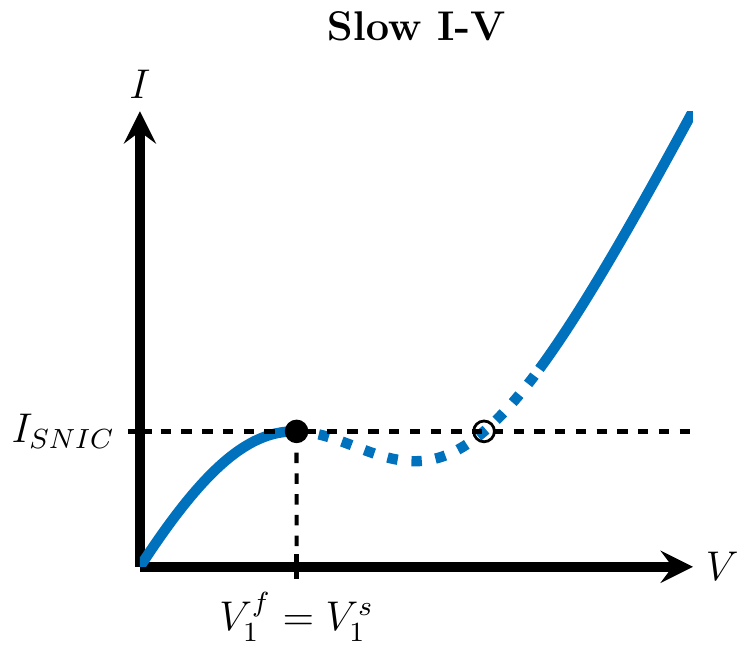}
\includegraphics[width=0.49\linewidth]{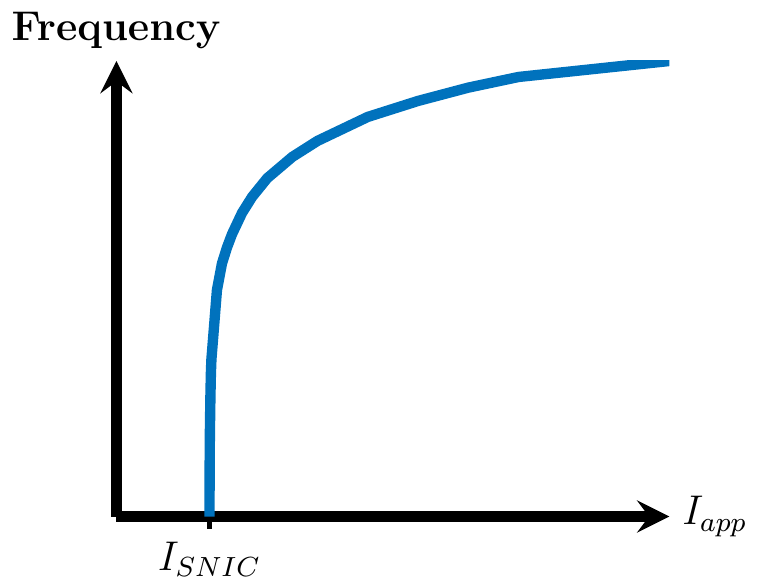}
\caption{Type 1 excitable neuron. Left: In order to generate oscillations with arbitrarily low frequency, it is necessary for the slow I-V curve to be non-monotonic, so that for $I_{app} = I_{SNIC}$ the system undergoes a saddle-node on invariant circle bifurcation. The thresholds are set so that the rest/spike bistability is lost, i.e. $V_1^f = V_1^s$. Right: The signature of the saddle-node on invariant circle bifurcation is the frequency of the oscillation tending to zero for $I_{app}$ close to $I_{SNIC}$.}
\label{fig:type1}
\end{figure}

The balance between $I_s^-$ and $I_s^+$ can also be used to control the properties of a purely spiking circuit, as this balance is central to controlling the spiking frequency of a neuron in the low frequency range \cite{drion_ion_2015}. Neurons that can spike at arbitrarily low frequency are referred to as Type I excitable neurons. Fig. \ref{fig:type1} illustrates the classical Type I neuron model governed by a SNIC bifurcation \cite{rinzel_analysis_1989}. The transition from Type II excitability in Fig. \ref{fig:excitable_frequency} to Type I in Fig. \ref{fig:type1} is achieved by shaping the slow I-V curve around its transition from monotone to ``N-shape'', so that $V_1^f$ and $V_1^s$ coincide. In the language of singularity theory, this transition is governed by a hysteresis singularity \cite{franci_modeling_2014}. In our circuit, shaping an I-V curve around a hysteresis singularity is achieved by balancing a positive and a negative conductance element. This robust regulatory mechanism is central to neuromodulation and can be repeated in any timescale.

\section{Fragile and rigid bursting mechanisms}

\label{sec:fragile_bursting}

Shaping the monotonicity properties of an I-V curve by balancing positive and negative conductance elements makes a bursting circuit robust and controllable \cite{franci_robust_2017}.
We will now briefly review two well-known bursting mechanisms \cite{rinzel_analysis_1989} that do not necessitate a negative conductance element $I_s^-$. Such models can burst but they lack the modulation properties described in the previous section. Both models have been prevalent in the bursting literature.

\begin{figure}[!t]
\centering
\includegraphics[width=1\linewidth]{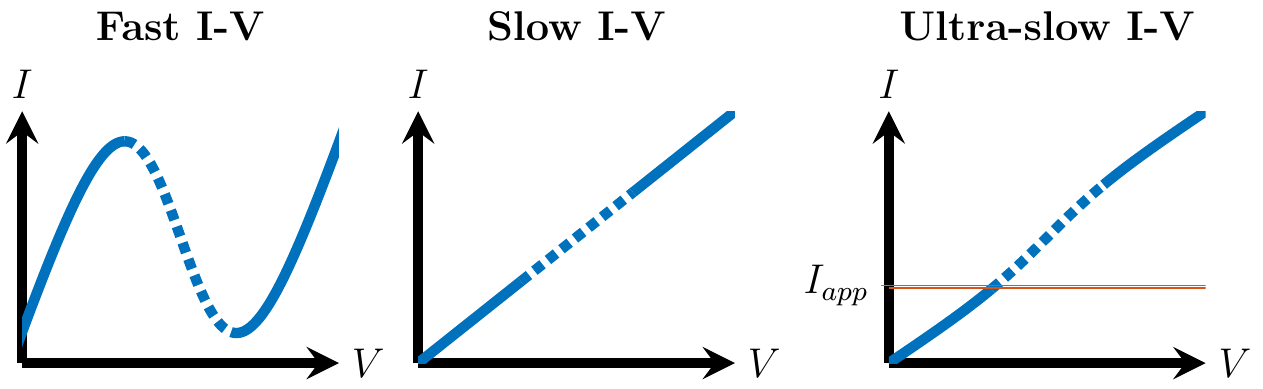}
\includegraphics[width=1\linewidth]{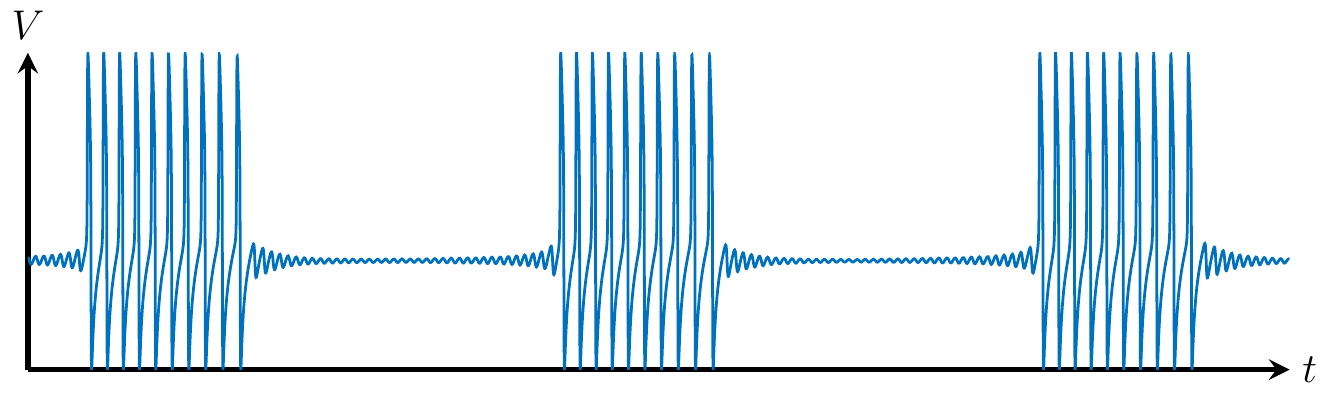}
\caption{Bursting through a subcritical Hopf bifurcation can be designed by adding an ultra-slow positive conductance element to a spiking circuit. This bursting mechanism is fragile to the time scale separation and only exists for a narrow range of applied current (depicted with the two horizontal lines in the ultra-slow I-V curve).}
\label{fig:bursting_hopf}
\end{figure}

The first bursting mechanism is illustrated in Fig. \ref{fig:bursting_hopf}. The fast and the slow I-V curves are the same as for the purely spiking circuit in Fig. \ref{fig:excitable_currents}. However, because the transition from resting to spiking is through a {\it subcritical} Hopf bifurcation, there exists a small range of applied current in which the system exhibits bistability between a fixed point and a stable limit cycle, separated by an unstable limit cycle. By introducing the ultra-slow positive conductance element that generates the ultra-slow oscillation between the two states, the system undergoes bursting oscillations.

These bursting oscillations only appear for a small range of values of $I_{app}$, and in addition to this, it is not possible to precisely determine this range from the ultra-slow I-V curve as before. The bistability range is also sensitive to timescale separation. In fact, it shrinks to zero as the timescale separation is increased \cite{franci_balance_2013}, unlike the slow bistability discussed in Fig. \ref{fig:slow_bistability}, which is robust to an increased timescale separation.

\begin{figure}[!t]
\centering
\includegraphics[width=1\linewidth]{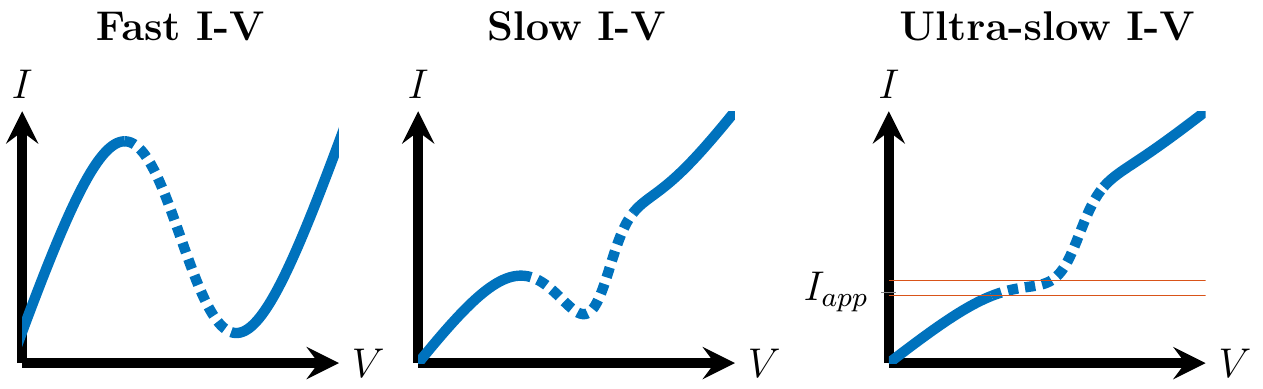}
\includegraphics[width=1\linewidth]{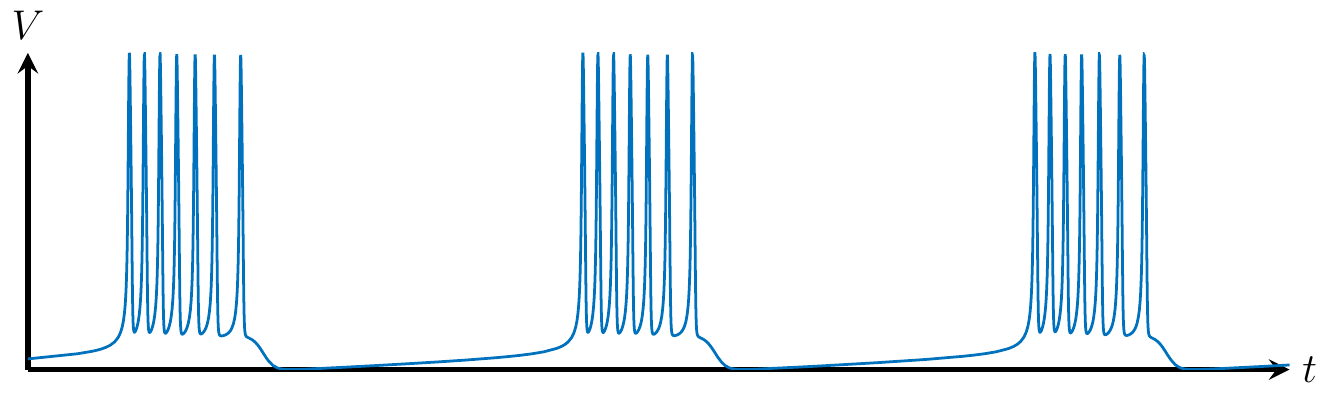}
\caption{Generating bursting from a Type I neuron. For a small timescale separation between fast and slow, the system can have bistability between resting and spiking, and adding ultra-slow positive conductance element turns it into a bursting circuit. The range of applied currents for which the system undergoes bursting oscillations is depicted with the two horizontal lines in the ultra-slow I-V curve.}
\label{fig:bursting_ml}
\end{figure}

The second bursting mechanism is illustrated in Fig. \ref{fig:bursting_ml}. It is achieved by adding the ultra-slow positive conductance element to a Type I neuron. We can construct a Type I neuron without the use of a slow negative conductance element $I_s^-$ by decreasing the linear range of the positive conductance element $I_s^+$ compared to the fast negative conductance element $I_f^-$, i.e. by having:
\begin{equation}
\label{eq:slow_ml}
I_s^{+} = \alpha_s^+ \tanh (\beta_s^+ (V_s - \delta_s^+)), \quad \beta_s^+ > 1
\end{equation}
Although a Type I neuron is monostable (Fig. \ref{fig:type1}), it can be turned into a bistable system if the fast and slow timescales are no longer separated. Such a construction is very sensitive to the particular choice of timescales, and making the slow timescale slower than approximately $2 \max(\tau_v,\tau_f)$ destroys the bistability. Bursting achieved in this way is shown in Fig. \ref{fig:bursting_ml}.

The bursting mechanisms illustrated in Fig.\ref{fig:bursting_hopf} and Fig. \ref{fig:bursting_ml} are not only fragile to parameter uncertainty. They are also rigid in the sense of completely lacking the tuning properties shown in Figs. \ref{fig:bursting}-\ref{fig:pitch}; in the first case, the system necessarily undergoes elliptic-type bursts, while the second undergoes plateau bursting oscillations, for which it is not possible to precisely control the height of the plateau, or the size of the fast spikes. The bursting synthesis in the previous sections is in sharp contrast with those mechanisms: the choice of timescales is inessential and all tuning properties can be directly deduced and designed by shaping the I-V curves of the circuit. These alternative constructions underline the value of a bursting circuit realized as the interconnection of both \textit{fast} and \textit{slow} excitability components in order to fully capture the tuning and robustness properties of biological neurons.

\section{Circuit implementation}

\subsection{Implementation of a localized conductance element}

\begin{figure}[!t]
\centering
\includegraphics[width=1\linewidth]{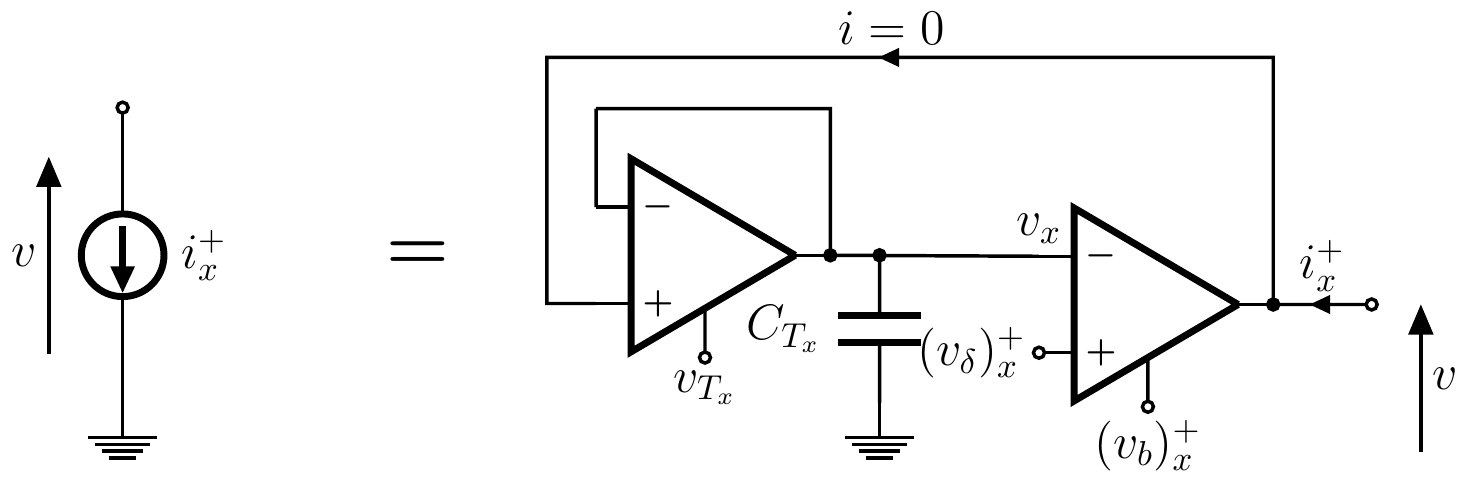}
\caption{Implementation of a single localized conductance element. The first transconductance amplifier and a capacitor form a non-linear first-order filter, whose output is the filtered voltage $v_x$. This is then the input to the second transconductance amplifier which forms the output current $i_x^+$. The element in this case is positive conductance (for a negative conductance element the inputs to the second amplifier are swapped).}
\label{fig:filter_amplifier}
\end{figure}

\begin{figure}[!t]
\centering
\includegraphics[width=1\linewidth]{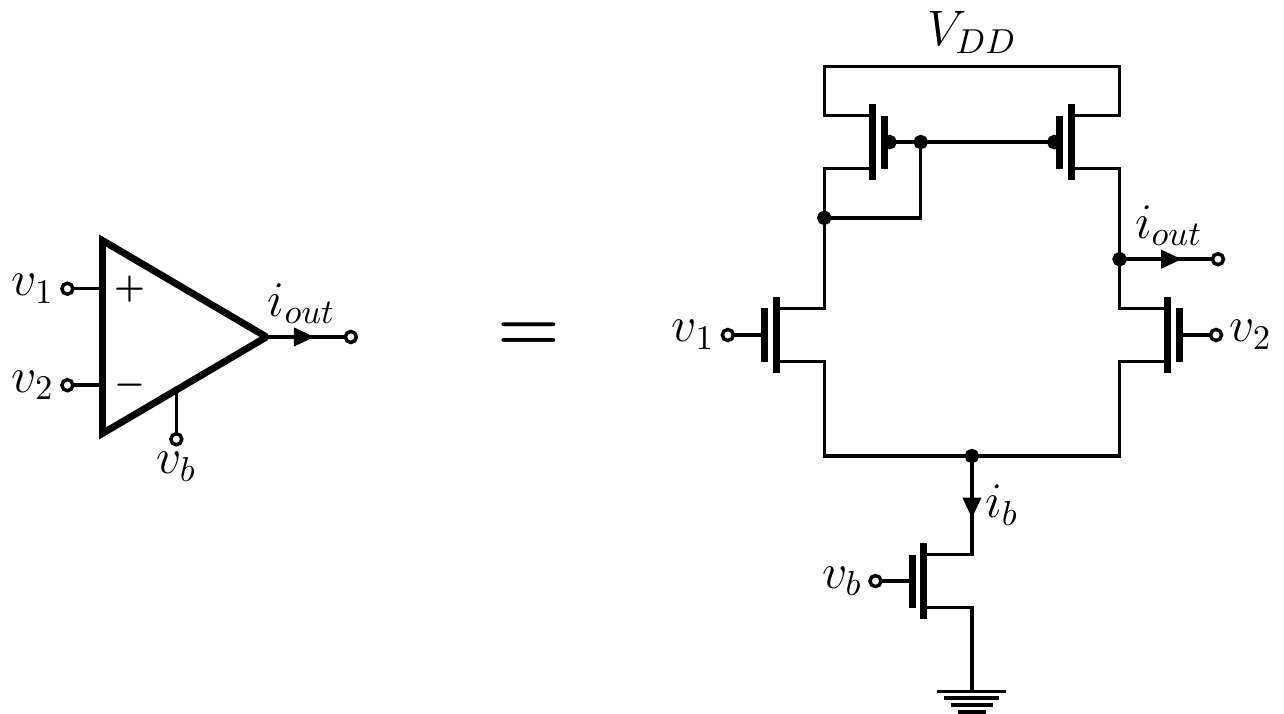}
\caption{A weak inversion MOSFET transconductance amplifier used as the circuit primitive. The circuit realizes a hyperbolic tangent mapping from the differential voltage input to the current output. The gain of the function is determined by the current flowing through the base transistor ($i_b$), controlled by its base voltage ($v_b$) that acts as the additional amplifier input.}
\label{fig:mosfet_transconductance}
\end{figure}

We propose a simple hardware implementation of an element that satisfies equation \eqref{eq:current_element} and has the I-V characteristic in \eqref{eq:localized_conductance}, as shown in Fig. \ref{fig:filter_amplifier}. We use a MOSFET-based transconductance amplifier operating in the weak inversion regime as the circuit primitive \cite{mead_analog_1989}. For our application, the simplest implementation of a transconductance amplifier is considered (Fig. \ref{fig:mosfet_transconductance}). It has the following input-output relationship:
\begin{equation}
\label{eq:transconductance_amp}
i_{out} = i_b \tanh \bigg( \kappa \frac{v_1 - v_2}{2 v_T} \bigg),
\end{equation}
where $i_b$ is the controlled current that sets the gain, $\kappa \in (0, 1)$ is a process-dependent variable, and $v_T \approx \SI{25}{\milli \volt}$ is the thermal voltage. Due to its well-defined hyperbolic tangent function and the adjustable gain, it is a versatile localized conductance element. The one control parameter, current $i_b$, is set by the bottom transistor (Fig. \ref{fig:mosfet_transconductance}, right) whose gate voltage is the third input on the block diagram representation (Fig. \ref{fig:mosfet_transconductance}, left). As the transistor is saturated in the weak inversion regime, the relationship between its gate voltage and base current is 
\begin{equation}
i_b = i_0 e^{\kappa v_b / v_T},
\end{equation}
where $i_0$ is the zero-bias current. Therefore, there is an exponential relationship between the gain of the element and its base voltage.

The conductance element has the following characteristic:
\begin{subequations}
\begin{align}
i_x^{\pm} &= \pm (i_b)^{\pm}_x \tanh \bigg( \kappa \frac{v_x - (v_{\delta})^{\pm}_x}{2 v_T} \bigg) \\
C_{T_x} \dot{v}_{x} &= i_{T_x} \tanh \bigg( \kappa \frac{v - v_{x}}{2v_T}          \bigg)
\end{align}
\end{subequations}
Its effective timescale is determined by the capacitance ($C_{T_x}$) and the base current ($i_{T_x}$), so that
\begin{equation}
T_x = (2 v_T C_{T_x}) / (\kappa i_{T_x})
\end{equation}

The current source acts as a positive conductance when the filtered voltage is connected to the negative input terminal, as in Fig. \ref{fig:filter_amplifier}. Instead, it acts as a negative conductance element if it is connected to the positive input terminal. The other terminal then defines the voltage offset, by default set to the middle of the voltage rails.

We map the dimensionless parameters from previous sections directly to the circuit parameters in the following way:
\begin{subequations}
\label{eq:parameter_mapping}
\begin{align}
(i_b)^{\pm}_x &= \alpha^{\pm}_x \bigg(\frac{2 G v_T}{\kappa} \bigg)\\
(v_{\delta})^{\pm}_x &= \delta^{\pm}_x \bigg(\frac{2 v_T}{\kappa} \bigg) \\
i_{app} &= I_{app} \bigg(\frac{2 G v_T}{\kappa} \bigg),
\end{align}
\end{subequations}
with $G$ being the conductance of the passive element around equilibrium, i.e.:
\begin{equation}
G = \frac{di_p(v)}{dv}\bigg|_{v=v_e}
\end{equation}

This conductance then defines the time constant of the voltage equation as in \eqref{eq:membrane_time_constant}.

\subsection{SPICE simulation}

In order to test the proposed architecture of the paper, we have simulated the circuit in SPICE environment, using the BSIM3 MOSFET model with TSMC \SI{0.35}{\micro\meter} process parameters and a \SI{3.3}{\volt} voltage supply.

Each current source element of the bursting circuit from Fig. \ref{fig:bursting_circuit} is realized as described in the previous section. The passive element is implemented by using a transconductance amplifier with an increased linear range. We linearize the element by using four diode-connected NMOS transistors to provide source-degeneration. Note that since currents $i_s^-$ and $i_s^+$ both act on the slow timescale, only one filter is necessary. 

All capacitors were chosen to have the same capacitance $C = \SI{100}{\pico \farad}$, and $G$ was set so that the period of the oscillation is of the order of seconds. We achieve this by setting:
\begin{equation}
T_{us} = \SI{1}{\second} = 50 \: T_s = 50^2 \: T_v
\end{equation}

\begin{figure}[!t]
\centering
\includegraphics[width=1\linewidth]{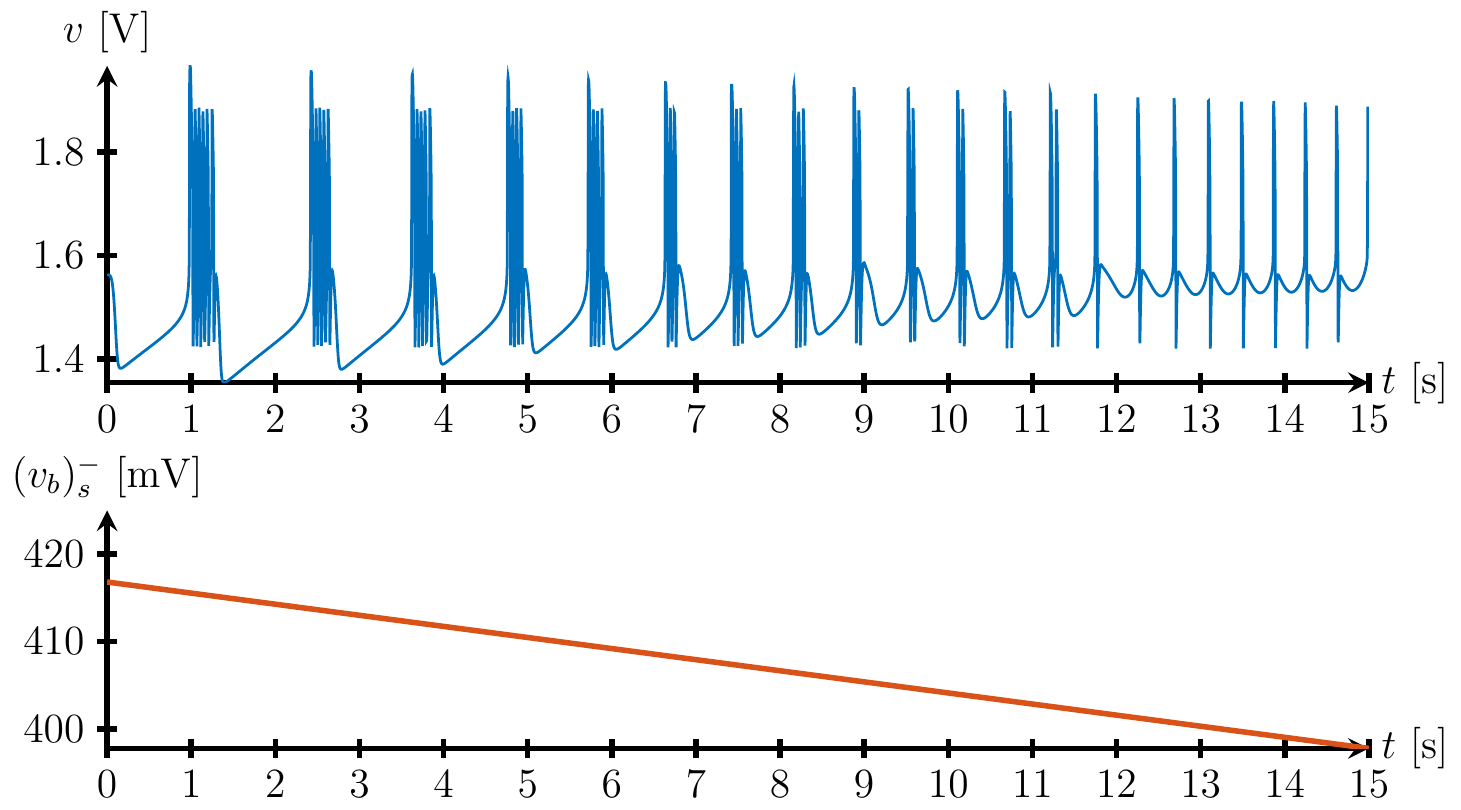}
\includegraphics[width=1\linewidth]{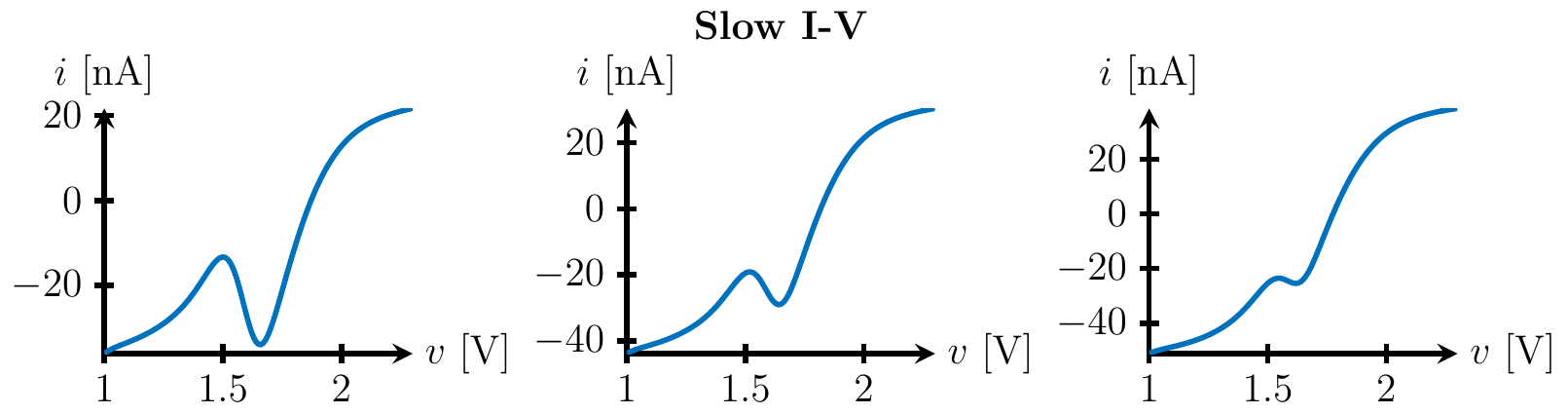}
\caption{Transition between bursting and spiking modes in the MOSFET circuit simulation. The gain of the slow negative conductance element is controlled through the corresponding base voltage that modifies the gain of the current element $i_s^-$. Decreasing $i_s^-$ reduced the negative conductance region of the slow I-V curve, changing the behavior from bursting to spiking.}
\label{fig:bursting_circuit_transition}
\end{figure}

The input transistors and the source degeneration transistors were chosen to have the minimal size, so that their width and length were set to $W = \SI{0.6}{\micro\meter}$ and $L = \SI{0.4}{\micro \meter}$. The bias transistors and the current mirror transistors were made larger in order to minimize the channel length modulation effect and improve matching, so that a more precise $\tanh$ current-voltage relationship is obtained. Their size was chosen to be $W = \SI{2.4}{\micro\meter}$ and $L = \SI{1.6}{\micro \meter}$. The total area taken by the transistors is \SI{84.96}{\micro\meter\squared}.

By using the relationships given in \eqref{eq:parameter_mapping} we can set the parameters of the circuit to replicate any behavior demonstrated in the previous sections. As an example, we concentrate on the transition from bursting to spiking from Section \ref{section:bursting_spiking} (Fig. \ref{fig:pitch}). We recreate this transition in Fig. \ref{fig:bursting_circuit_transition}. The transition is controlled by the parameter $\alpha_s^-$, which in the circuit corresponds to the base current $(i_b)^{-}_s$ of the $i_s^-$ localized conductance element. Due to the nonlinearity of the passive element, the bursts have a slightly larger amplitude of the slow oscillation for high values of $(v_b)_s^-$ as seen in the figure, but we observe exactly the same transition as the negative conductance region of the slow I-V curve is modulated. The correspondence between the simulations can be improved by choosing a better linearization technique, but this does not affect the excitability properties of the circuit.

The simulated power consumption of the circuit is \SI{0.77}{\micro \watt} in the bursting regime.

\subsection{Robustness of I-V curve shaping}

\begin{figure}[!t]
\centering
\includegraphics[width=1\linewidth]{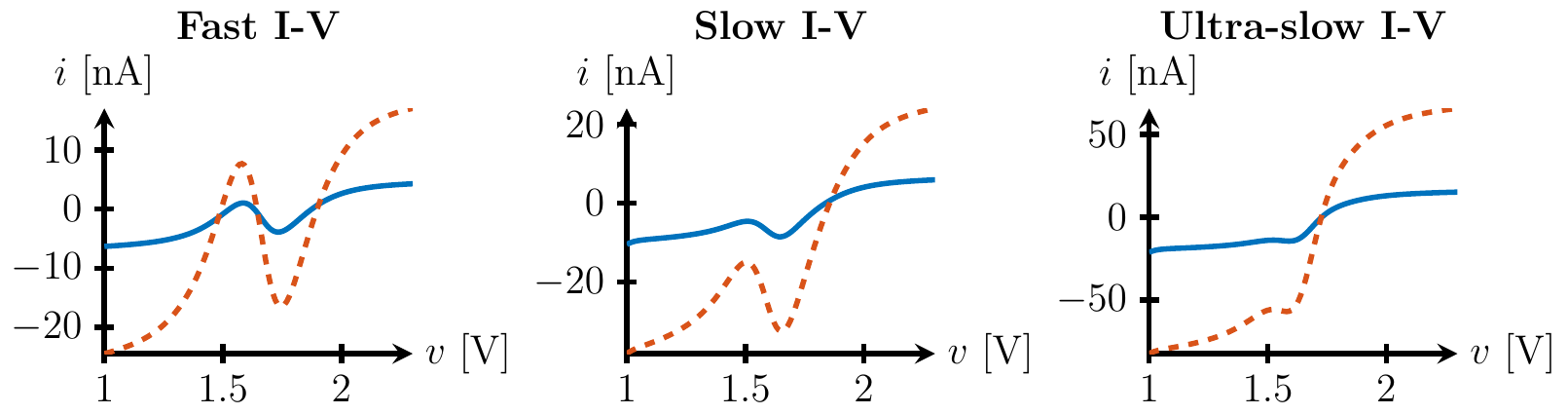}
\includegraphics[width=1\linewidth]{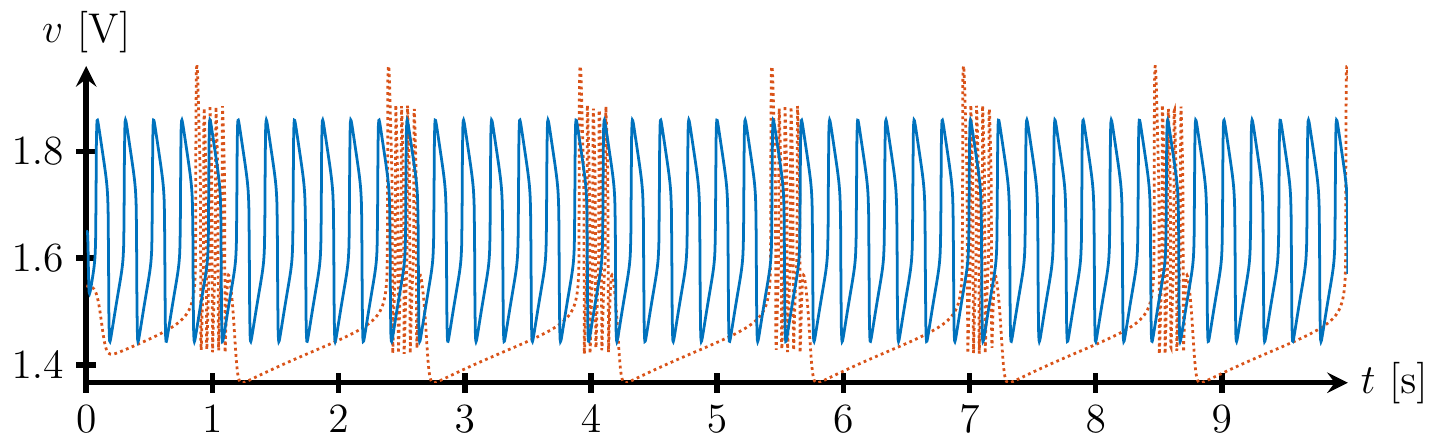}

\vspace{0.5cm}

\includegraphics[width=1\linewidth]{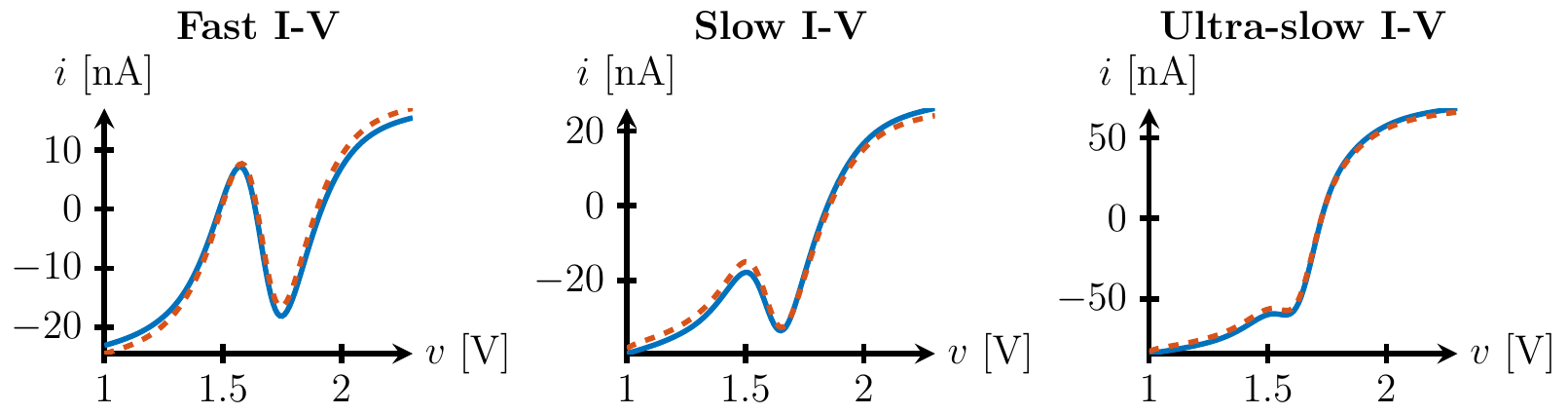}
\includegraphics[width=1\linewidth]{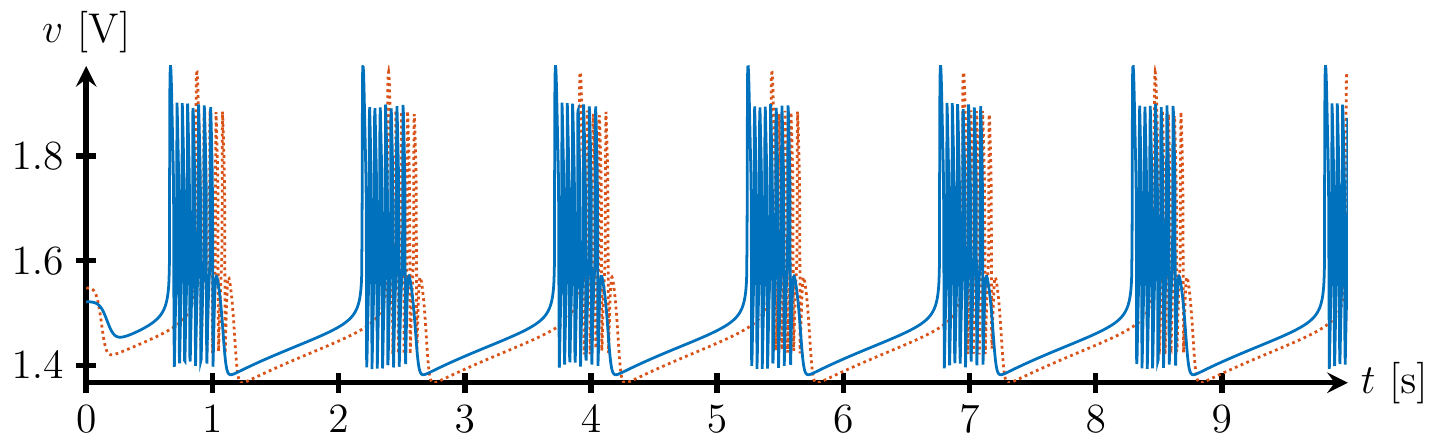}
\caption{Compensating for process variation. The element gains of a circuit instance differing from the nominal can be adjusted so that the I-V curves are restored and the behavior is kept the same. Top: Due to process variation, the I-V curves of the circuit (blue, solid) are distorted compared to the nominal (orange, dashed), and the bursting behavior is lost. Bottom: After compensation, the circuit's I-V curves closely match the nominal I-V curves, and the bursting behavior is restored.}
\label{fig:bursting_circuit_transition_var}
\end{figure}

Due to manufacturing uncertainty, the parameters of the circuit will vary from the idealistic conditions of the previous section. To this extent, we would like to stress two important characteristics of our proposed architecture:
\begin{enumerate}
\item Maintaining the I-V curves of the circuit keeps the behavior intact.
\item The circuit's I-V curves can be fully controlled through the gain and offset voltages of the localized conductance elements.
\end{enumerate}

The first point effectively means that the underlying circuit structure is inessential: as long as the input-output characteristic of the circuit consists of a specific set of I-V curves, its behavior is well-defined. Robustness of the circuit behavior therefore boils down to the robustness of its I-V curves.

The second point stresses that the internal parameters of the circuit can be readjusted so that the set of I-V curves is kept constant. This means that the variability in the components can be compensated for by tuning the control voltages of the circuit elements.

To show this, we have investigated how process variability affects the I-V relationships of the circuit elements by varying the following process parameters: threshold voltage, surface mobility at the nominal temperature, gate oxide thickness, and transistor width and length offset parameters. For each parameter, the variability was modeled as a Gaussian distribution around the nominal value with the standard deviation at $10\%$ of the nominal value.

The main effect of these process variations on the I-V characteristics of the elements was found to be the variation in the gains of the localized conductance elements, while the shape of the $\tanh$ relationships remained largely unchanged. As a result, such variations can be compensated for by readjusting the base voltages controlling the element gains ($v_b$ in Fig. \ref{fig:mosfet_transconductance}). We show this on a random instance of the circuit in Fig. \ref{fig:bursting_circuit_transition_var}: for nominal values the bursting behavior is lost, but by rescaling the element gains the I-V curves are restored, and therefore, the nominal behavior. Such variability can be compensated for if the variation in the element's base currents is between $0.3$ and  $3$ times its nominal value, dictated by the voltage limits of the subthreshold operating region of MOSFETs. The local transistor mismatches introduce voltage offsets in the $\tanh$ I-V relationships \cite{simoni_multiconductance_2004}, and such variations are compensated for through the control of offset voltages of the circuit elements ($v1$ or $v2$ in Fig. \ref{fig:mosfet_transconductance}).

\begin{figure}[!t]
\centering
\includegraphics[width=1\linewidth]{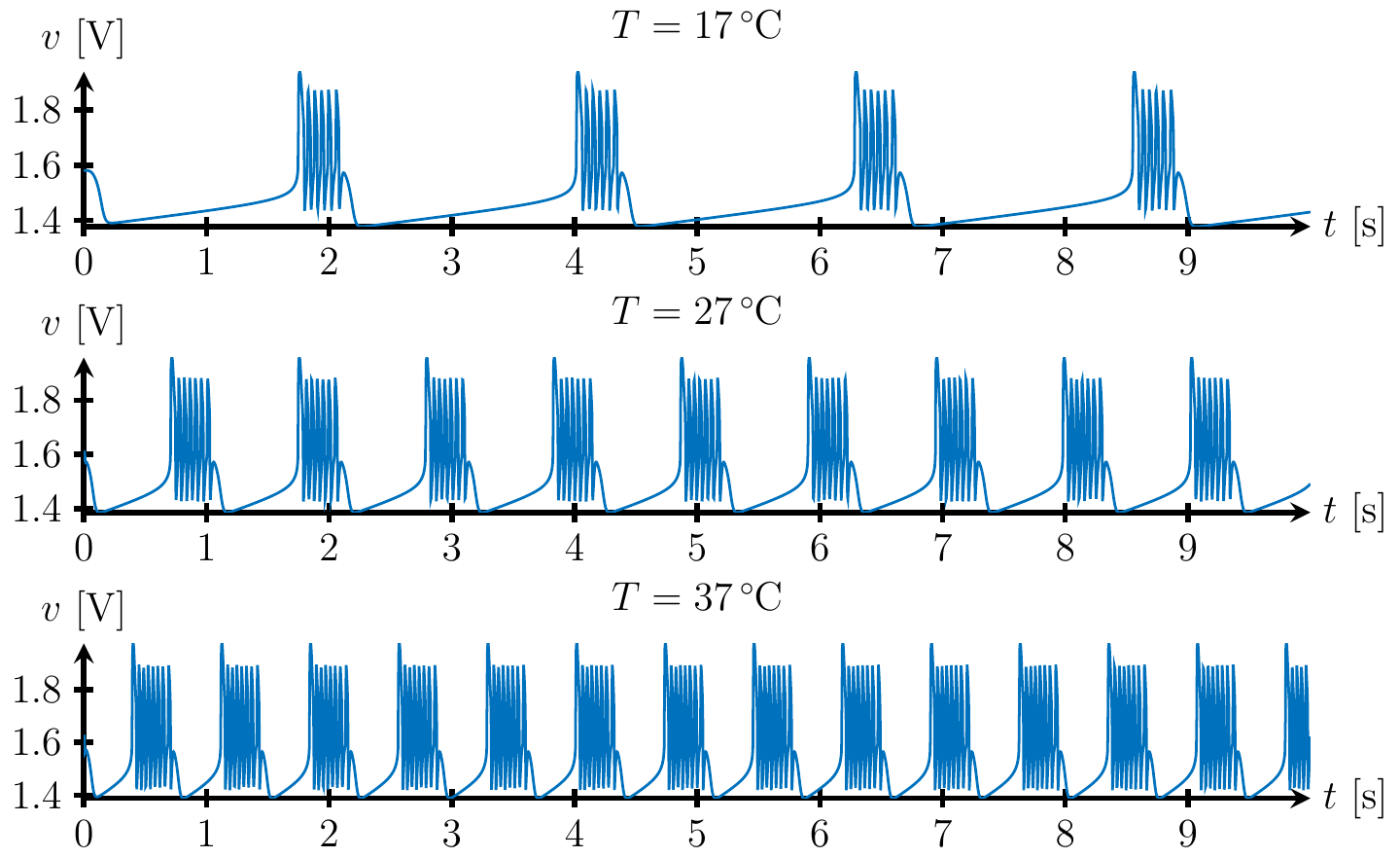}
\caption{Temperature dependence of the circuit. As temperature is increased, the circuit maintains the bursting oscillation with increasing interburst frequency, mimicking the behavior of biological neurons.}
\label{fig:bursting_temperature}
\end{figure}

We also consider how temperature variations affect the behavior of the circuit. In Fig. \ref{fig:bursting_temperature} we can see that the circuit maintains the bursting behavior for changing temperature, while the interburst frequency increases with increasing temperature. Such a dependence on temperature is in correspondence with the common behavior of biological neurons \cite{rinberg_effects_2013}.

In order to account for the variability in circuit components as well as temperature variations, additional compensation techniques can be considered that would allow the circuit to automatically maintain its behavior irrespective of the changing conditions (see e.g. \cite{simoni_adaptation_1999}). As we have discussed in this section, the aim of such techniques would be to maintain the circuit's input-output properties in the form of its I-V curves, so that the behavior is subsequently preserved.

\section{Conclusion}

We have presented a novel methodology for realizing neuromorphic circuits with robustness and neuromodulation properties reminiscent of those observed in physiological neurons. The methodology is based on shaping the circuit's input-output I-V curves in several timescales, which gives a robust and flexible approach that relies on I-V curve shaping. It departs from the standard methodology of achieving a specific behavior by specific parameter tuning, which often leads to circuits that lack modulation capabilities. As a proof of concept, we have proposed a simple circuit implementation using MOSFET transconductance amplifiers operating in the weak inversion regime.

While the proposed methodology is qualitative, all behaviors analyzed in this paper have low-dimensional state-space model realizations amenable to an exhaustive and rigorous bifurcation analysis. This is because the I-V curve shaping methodology is closely related to the analysis of those dynamical models by singularity theory. All the attractors studied in the present paper are organized by the same algebraic singularity. The reader interested in a detailed analysis is referred to \cite{franci_modeling_2014}.

The main perspective of our approach is twofold. By retaining the basic interconnection structure of biophysical models, the circuit can be utilized to better understand the modulation principles of real neurons in experimental setups. Recent work \cite{drion_dynamic_2015} has shown how the transitions in neural behavior can be predicted and traced by accumulating the effects of ionic conductances in several relevant timescales. In turn, our localized conductance elements capture these collective effects, so that by designing different behaviors in hardware, we are able to predict which neuromodulators would need to be utilized to capture the same transitions in experiments.

Secondly, by retaining the relevant tuning mechanisms in hardware, we aim to study how the transitions between different neural behaviors are utilized in biology. We are hopeful this will lead to novel signal processing paradigms, where the spatio-temporal characteristics of the neural waveforms can be effectively used to capture the sensory information in more efficient ways, mimicking the operation and structure of biological neural architectures. The relevance of such mechanisms is suggested in studies such as \cite{krahe_burst_2004}. Robust neuromodulation is also at the essence of the remarkable adaptability of neural central pattern generators (CPGs) that generate movement in animals. Such mechanisms have been extensively investigated in robotics in order to build autonomous robots that adapt efficiently to changing environmental conditions. However, a solid theoretical understanding for the robust and adaptive hardware of CPGs  is still  lacking \cite{ijspeert_central_2008,yu_survey_2014}. The I-V curve shaping methodology offers a promising avenue for building tunable CPG circuits. For instance, we are confident that the circuit proposed in the present paper provides a building block for a CPG with the neuromodulation capabilities described in the recent paper  \cite{drion_cellular_2018}.

Our feasibility study of a simplistic MOSFET circuit operating in the weak inversion regime built on the I-V curve shaping principles shows the potential of our approach to introduce neuromodulation principles in neuromorphic hardware. The hardware implementation of the proposed circuit will be further discussed in a future study.

\begin{table*} [!t]
\caption{Parameter values for simulations in sections \ref{sec:excitable} - \ref{sec:fragile_bursting}}
\centering
\label{table:parameters}
\centering
\begin{tabular} { |c|c|c|c|c|c|c|c|c|c| }
\hline
{} & $\alpha_f^-$ & $\delta_f^-$ & $\alpha_s^+$ & $\delta_s^+$ & $\alpha_s^-$ & $\delta_s^-$ & $\alpha_{us}^+$ & $\delta_{us}^+$ & $I_{app}$ \\
\hline
Fig. \ref{fig:excitable_currents} & 2 & 0 & 2 & 0 & / & / & / & / & 0, -1 \\
\hline
Fig. \ref{fig:excitable_amp} & 2.5, 2, 1.5 & 0 & 2.5, 2, 1.5 & 0 & / & / & / & / & 0 \\
\hline
Fig. \ref{fig:excitable_frequency} (left) & 2 & 0 & [2,4] & 0 & / & / & / & / & -0.8 \\
\hline
Fig. \ref{fig:excitable_frequency} (right) & 2 & 0 & 2 & 0 & / & / & / & / & [-1,-0.2] \\
\hline
Fig. \ref{fig:slow_spiking} & / & / & / & / & 1.5 & -0.88 & 2 & 0 & -2, -2.6 \\
\hline
Fig. \ref{fig:bursting} & 2 & 0 & 2 & 0 & 1.5 & -0.88 & 2 & 0 & -1, -2, -2.6 \\
\hline
Fig. \ref{fig:bursting_amplitude} (top) & 2 & 0 & 2 & 0 & 1.5 & -1.5 & 1.5 & -1.5 & -2 \\
\hline
Fig. \ref{fig:bursting_amplitude} (bottom) & 2 & 0 & 2 & 0 & 1.3 & -1 & 1.3 & -1 & -1 \\
\hline
Fig. \ref{fig:bursting_frequency} (left) & 2 & 0 & 2, 2.2, 2.6 & 0 & 1.5, 1.65, 1.95 & -0.88 & 1.5 & -0.88 & -1.3 \\
\hline
Fig. \ref{fig:bursting_frequency} (right) & 2 & 0 & 2 & 0 & 1.5 & -0.88 & 1.5, 2.5, 3.5 & -0.88 & -1.3 \\
\hline
Fig. \ref{fig:pitch} & 2 & 0 & 2 & 0 & [0.8,1.6] & -0.88 & 2 & 0 & -2.2 \\
\hline
Fig. \ref{fig:type1} & 2 & 0 & 2 & 0 & 1.2 & -0.45 & / & / & [-0.4,-0.3] \\
\hline
Fig. \ref{fig:bursting_hopf} & 2 & 0 & 2 & 0 & / & / & 0.5 & 0 & -1.2 \\
\hline
\end{tabular}
\end{table*}

\appendices
\section{Transcritical bifurcation in the bursting model}
\label{appendix:a}
We connect the I-V curve analysis presented in this paper with previous mathematical analysis of a simplified bursting model \cite{franci_modeling_2014}. We concentrate on the fast-slow dynamics of the model that has the rest/spike bistability property. The simplified model is the following:
\begin{subequations}
\label{eq:siam_model}
\begin{align}
\dot{x} &= -x^3 + \beta x - (y + \lambda)^2 + \alpha  \\
\dot{y} &= \varepsilon (x - y)
\end{align}
\end{subequations}
with $\varepsilon \ll 1$ so that $x$ is the fast voltage variable, and $y$ is the slow variable.

For a range of parameters, this model experiences bistability between a stable rest state and a stable limit cycle. This is exhibited in the fast-slow phase portrait as a mirrored hysteresis fast nullcline; in the limit of timescale separation, i.e. when $\varepsilon = 0$, the bistability is lost when the two hysteresis branches meet at a transcritical singularity.

The conditions on the I-V curves stated in Fig. \ref{fig:slow_bistability} directly relate our bursting circuit to this model. We will consider the case where $\tau_f = 0$, so that the fast-slow system consists of two state variables, $V$ and $V_s$. Firstly, we look at the requirements for the fast I-V curve:
\begin{equation}
I_p(V) + I_f^-(V) = V - \alpha_f^- \tanh(V)
\end{equation}
Without loss of generality, we assume $\delta_f^- = 0$. In order for the fast I-V curve to be ``N-shaped'', we require $\alpha_f > 1$; when this is the case, the fast I-V curve is locally equivalent to the instantaneous term in Eq. \eqref{eq:siam_model} $x^3 - \beta x$, for $\beta \neq 0$, which can be easily verified by checking the first three derivatives around the points $V = 0$ and $x = 0$, respectively.

Because the slow I-V curve is obtained by adding the slow conductance characteristic $I_s(V) = I_s^+(V) + I_s^-(V)$ to the fast I-V curve, we can infer the properties of the slow conductance from the I-V curve conditions. The conditions that the slow I-V curve is ``N-shaped'' and that $V_1^s < V_1^f$ means that for $V \in (V_1^s,V_1^f)$, the fast I-V curve has a positive slope and the slow I-V curve has a negative slope, and vice-versa for $V \in (V_2^s,V_2^f)$. This means that the slow negative conductance element necessarily has a negative slope in the first region, and a positive slope in the second region, therefore having a local minimum for some $V_*^s \in (V_1^f,V_2^s)$. The characteristic is therefore locally quadratic around the point $V_*^s$, so that the I-V characteristic of the slow conductance elements is locally equivalent to the quadratic slow term in Eq. \eqref{eq:siam_model}.

This allows us to find the point of the transcritical singularity by having the following conditions on the derivatives of the fast nullcline around the point ($V_{tr}$,$V_{tr}^s$):

\begin{equation}
\label{eq:transcritical}
\frac{\partial \dot{V}}{\partial V} \bigg|_{V=V_{tr},V_s=V_{tr}^s} = \frac{\partial \dot{V}}{\partial V_s} \bigg|_{V=V_{tr},V_s=V_{tr}^s} = 0
\end{equation}

Solving \eqref{eq:transcritical}, we obtain:

\begin{subequations}
\label{eq:transcritical2}
\begin{align}
\frac{d}{dV} \Bigg( I_p^+(V) + I_f^-(V) \Bigg)\Biggr|_{V = V_{tr}} &= 0 \\
\frac{d}{dV_s} \Bigg(  I_s^+(V_s) + I_s^-(V_s) \Bigg)\Biggr|_{V_s = V_{tr}^s} &= 0
\end{align}
\end{subequations}

Following from \eqref{eq:transcritical2}, we get:
\begin{equation}
(V_{tr}, V_{tr}^s) = (V_1^f,V_*^s)
\end{equation}
so that these points correspond to the maximum of the fast I-V curve, and the minimum of the slow conductance I-V characteristic, respectively. We find the corresponding $I_{app}$ by imposing that this point lies on the fast nullcline, so that finally:
\begin{equation}
I_2^s = I_p^+(V_1^f) + I_f^-(V_1^f) + I_s^+(V_*^s) + I_s^-(V_*^s)
\end{equation}


\section{Simulation parameters}
\label{appendix:b}
All simulations in Sections \ref{sec:excitable} - \ref{sec:fragile_bursting} were carried out in MATLAB.

Figs. \ref{fig:excitable_currents}, \ref{fig:excitable_amp}, \ref{fig:excitable_frequency}, \ref{fig:slow_spiking}, \ref{fig:bursting}, \ref{fig:bursting_amplitude}, \ref{fig:bursting_frequency}, \ref{fig:pitch}, \ref{fig:type1}, \ref{fig:bursting_hopf} use the model described in \eqref{eq:dimensionless}, \eqref{eq:localized_conductance} and \eqref{eq:passive}.

The parameters for each conductance element are given in Table \ref{table:parameters}, as well as the applied currents. Common parameters for these figures are the following:
\begin{align*}
\tau_f &= 0 \\
\tau_s &= 50 \\
\tau_{us} &= 2500
\end{align*}

In Fig. \ref{fig:bursting_ml} the slow timescale is modified to $\tau_s = 2$, and the model described in \eqref{eq:slow_ml} is used for the slow positive conductance. The parameters are the following: $\alpha_f^- = 2$, $\delta_f^- = 0$, $\alpha_s^+ = 1$, $\beta_s^+ = 3$, $\delta_s^+ = 0.5$, $\alpha_{us}^+ = 1$, $\delta_{us}^+ = 0$.


\section*{Acknowledgment}
The authors gratefully acknowledge numerous discussions with Alessio Franci and Guillaume Drion during the preparation of this paper.

\ifCLASSOPTIONcaptionsoff
  \newpage
\fi

\end{document}